\documentclass[a4paper,11pt]{article}
\usepackage{amsmath,amsfonts,amssymb,graphicx,epsfig,multirow}
 \graphicspath{{./eps/}}
\numberwithin{equation}{section}
\usepackage{cite}
\usepackage[usenames]{color}
\usepackage{pstricks}
\usepackage[backref=false]{hyperref}
\hypersetup{
colorlinks=true,
citecolor=red,
linkcolor=darkblue
}
\definecolor{darkblue}{rgb}{0,0,.8}
\definecolor{red}{rgb}{1,0,0}

\long\def\ignore#1{}
\definecolor{purple}{rgb}{1,0,1}
\definecolor{darkpurple}{rgb}{1,.2,1}
\definecolor{pink}{rgb}{1,.7,.7}
\definecolor{tile}{rgb}{1,.9,.7}

\newcommand{\nc}{\newcommand}
\nc\disp{\displaystyle}
\nc{\fh}{\hat{f}}
\nc{\negg}{\negthickspace}
\nc{\muh}{\hat{\mu}}
\nc{\nuh}{\hat{\nu}}
\nc{\spos}[2]{\makebox(0,0)[#1]{$\sm{#2}$}}
\nc{\sm}[1]{{\scriptstyle #1}}
\nc{\qbar}{\overline{q}}
\def\Tr{\mathop{\mbox{Tr}}}
\nc{\bib}{\bibitem}
\nc{\al}{\alpha}
\nc{\g}{\gamma}
\nc{\G}{\Gamma}
\nc{\D}{\Delta}
\nc{\eps}{\epsilon}
\nc{\la}{\lambda}
\nc{\La}{\Lambda}
\nc{\var}{\varphi}
\nc{\pa}{\partial}
\nc{\nn}{\nonumber \\ }
\nc{\hf}{\frac{1}{2}}
\nc{\dz}{\frac{dz}{2\pi i}}
\nc{\bin}[2]{\left(\!\!\!\begin{array}{c} {#1}\\ {#2} \end{array}\!\!\!\right)}
\nc{\be}{\begin{equation}}
\nc{\ee}{\end{equation}}
\nc{\bea}{\begin{eqnarray}}
\nc{\eea}{\end{eqnarray}}
\nc{\bra}[1]{\langle {#1}|}
\nc{\ket}[1]{|{#1}\rangle}
\nc{\ketw}[1]{({#1})^{\phantom{a}}_{{\cal W}}}
\nc{\chit}{\raisebox{0.25ex}{$\chi$}}
\nc{\chih}{\raisebox{0.25ex}{$\hat\chi$}}
\nc{\Db}{\mbox{\boldmath $D$}}
\nc{\Hb}{\mbox{\boldmath $H$}}
\nc{\calH}{{\cal H}}
\nc{\calR}{{\cal R}}
\nc{\calL}{{\cal L}}
\nc{\calV}{{\cal V}}
\nc{\Hc}{{\cal H}}
\nc{\Rc}{{\cal R}}
\nc{\Lc}{{\cal L}}
\nc{\Vc}{{\cal V}}
\nc{\Ib}{\mbox{\boldmath $I$}}
\nc{\qb}{\bar{q}}
\nc{\Ac}{\mathcal{A}}
\nc{\Bc}{\mathcal{B}}
\nc{\Cc}{\mathcal{C}}
\nc{\Dc}{\mathcal{D}}
\nc{\Ec}{\mathcal{E}}
\nc{\Gc}{\mathcal{G}}
\nc{\Ic}{\mathcal{I}}
\nc{\Jc}{\mathcal{J}}
\nc{\Oc}{\mathcal{O}}
\nc{\Pc}{\mathcal{P}}
\nc{\Sc}{\mathcal{S}}
\nc{\Tc}{\mathcal{T}}
\nc{\Wc}{\mathcal{W}}
\nc{\Xc}{\mathcal{X}}
\nc{\Yc}{\mathcal{Y}}
\nc{\Zc}{\mathcal{Z}}
\nc{\fus}{\mbox{}\,\hat\otimes\,\mbox{}}
\nc{\Pch}{\hat{\Pc}}
\nc{\Rch}{\hat{\Rc}}
\nc{\Dh}{\hat{\Delta}}
\nc{\rh}{\hat{r}}
\nc{\sh}{\hat{s}}
\nc{\taub}{\bar{\tau}}
\nc{\Jcb}{\Jc_{\mathrm{b}}}
\nc{\rtt}{\mathtt{r}}
\nc{\stt}{\mathtt{s}}
\nc{\cosR}{\cos\frac{\pi p'rr'}{p}}
\nc{\cosS}{\cos\frac{\pi pss'}{p'}}
\nc{\sinR}{\sin\frac{\pi p'rr'}{p}}
\nc{\sinS}{\sin\frac{\pi pss'}{p'}}
\def\vvdots{\mathinner{\mkern1mu\raise1pt\vbox{\kern7pt\hbox{.}}\mkern2mu
  \raise4pt\hbox{.}\mkern2mu\raise7pt\hbox{.}\mkern1mu}}
\nc{\gauss}[2]{\big[\!\!\begin{array}{c} {#1}\\ {#2} \end{array}\!\!\big]}
\nc{\bgauss}[2]{\Big[\!\!\begin{array}{c} {#1}\\ {#2} \end{array}\!\!\Big]}
\nc{\sbin}[2]{\left\{\!\!\!\begin{array}{c} {#1}\\ {#2}
\end{array}\!\!\!\right\}}
\nc{\sbinlr}[2]{\Big\langle\!\!\begin{array}{c} {#1}\\ {#2}
\end{array}\!\!\Big\rangle}
\nc{\bino}[2]{\left(\!\!\begin{array}{c} {#1}\\ {#2} \end{array}\!\!\right)}
\def\half {\mbox{$\textstyle \frac{1}{2}$ }}
\def\vec#1{\mbox {\boldmath $#1$}}

\definecolor{lightblue}{rgb}{.61,.61,1}
\definecolor{midblue}{rgb}{.7,.7,1}
\definecolor{lightlightblue}{rgb}{.85,.85,1}
\definecolor{lightestblue}{rgb}{.96,.96,1}
\definecolor{lightpurple}{rgb}{1,.65,1}
\nc{\ch}{{\rm ch}}
\nc{\R}{{\cal R}}
\nc{\dkk}{\delta_{j,\{k,k'\}}^{(2)}}
\nc{\drr}{\delta_{j,\{r,r'\}}^{(2)}}
\nc{\ddkk}{\delta_{j,\{k,k'\}}^{(4)}}
\nc{\dddkk}{\delta_{j,\{k,k'\}}^{(8)}}
\nc{\dnn}{\delta_{j,\{n,n'\}}^{(2)}}
\nc{\ddnn}{\delta_{j,\{n,n'\}}^{(4)}}
\nc{\dddnn}{\delta_{j,\{n,n'\}}^{(8)}}

\definecolor{pink}{rgb}{1,.65,.65}

\thicklines

\def\faceD#1#2#3#4{\ \ 
\begin{pspicture}[shift=-.6](0,-.25)(1,1.25)
\pspolygon[linewidth=.5pt,fillstyle=solid,fillcolor=tile](0,0)(1,0)(1,1)(0,1)(0,0)
\rput[tr](0,0){\scriptsize $#1$}
\rput[tl](1,0){\scriptsize $#2$}
\rput[bl](1,1){\scriptsize $#3$}
\rput[br](0,1){\scriptsize $#4$}
\psline[linewidth=1pt,linestyle=dashed,linecolor=blue](0,0)(1,1)
\psline[linewidth=1pt,linestyle=dashed,linecolor=red](0,1)(1,0)
\end{pspicture}}

\def\faceF#1#2#3#4{\ \ 
\begin{pspicture}[shift=-.6](0,-.25)(1,1.25)
\pspolygon[linewidth=.5pt,fillstyle=solid,fillcolor=tile](0,0)(1,0)(1,1)(0,1)(0,0)
\rput[tr](0,0){\scriptsize $#1$}
\rput[tl](1,0){\scriptsize $#2$}
\rput[bl](1,1){\scriptsize $#3$}
\rput[br](0,1){\scriptsize $#4$}
\psline[linewidth=1pt,linestyle=dashed,linecolor=red](0,0)(1,1)
\psline[linewidth=1pt,linestyle=dashed,linecolor=blue](0,1)(1,0)
\end{pspicture}}

\def\faceX#1#2#3#4{\ \ 
\begin{pspicture}[shift=-.6](0,-.25)(1,1.25)
\pspolygon[linewidth=.5pt,fillstyle=solid,fillcolor=tile](0,0)(1,0)(1,1)(0,1)(0,0)
\rput[tr](0,0){\scriptsize $#1$}
\rput[tl](1,0){\scriptsize $#2$}
\rput[bl](1,1){\scriptsize $#3$}
\rput[br](0,1){\scriptsize $#4$}
\psline[linewidth=1pt,linestyle=dashed,linecolor=blue](0,0)(1,1)
\psline[linewidth=1pt,linestyle=dashed,linecolor=blue](0,1)(1,0)
\end{pspicture}}

\def\faceu#1#2#3#4#5{\ \ 
\begin{pspicture}[shift=-.6](0,-.25)(1,1.25)
\pspolygon[linewidth=.5pt,fillstyle=solid,fillcolor=lightlightblue](0,0)(1,0)(1,1)(0,1)(0,0)
\rput[tr](0,0){\scriptsize $#1$}
\rput[tl](1,0){\scriptsize $#2$}
\rput[bl](1,1){\scriptsize $#3$}
\rput[br](0,1){\scriptsize $#4$}
\rput(.5,.5){\small $#5$}
\psarc[linewidth=1pt,linecolor=red](0,0){.15}{0}{90}
\end{pspicture}}

\def\diam#1#2#3#4#5{\ \ 
\begin{pspicture}[shift=-.65](0,0)(1.53,1.5)
\pspolygon[linewidth=.5pt,fillstyle=solid,fillcolor=lightlightblue](.75,0)(1.5,.75)(.75,1.5)(0,.75)(.75,0)
\rput[t](.75,0){\scriptsize $#1$}
\rput[l](1.5,.75){\scriptsize $#2$}
\rput[b](.75,1.5){\scriptsize $#3$}
\rput[r](0,.75){\scriptsize $#4$}
\rput(.75,.75){\small $#5$}
\psarc[linewidth=1pt,linecolor=red](.75,0){.15}{45}{135}
\end{pspicture}}

\def\tile#1#2#3#4{\ \ 
\begin{pspicture}[shift=-.65](0,0)(1.5,1.5)
\pspolygon[linewidth=.5pt,fillstyle=solid,fillcolor=tile](.75,0)(1.5,.75)(.75,1.5)(0,.75)(.75,0)
\rput[t](.75,0){\scriptsize $#1$}
\rput[l](1.5,.75){\scriptsize $#2$}
\rput[b](.75,1.5){\scriptsize $#3$}
\rput[r](0,.75){\scriptsize $#4$}
\end{pspicture}}

\def\shifttile#1#2#3#4{\ \ 
\begin{pspicture}[shift=-.65](0,0)(1,1)
\pspolygon[linewidth=.5pt,fillstyle=solid,fillcolor=tile](0,0)(1,0)(1,1)(0,1)(0,0)
\psline[linewidth=1pt,linecolor=blue,linestyle=dashed](0,0)(1,1)
\rput[B](0,-.2){\scriptsize $#1$}
\rput[B](1,-.2){\scriptsize $#2$}
\rput[B](1,1.15){\scriptsize $#3$}
\rput[B](0,1.15){\scriptsize $#4$}
\end{pspicture}}

\def\tilea{\;
\begin{pspicture}[shift=-.65](0,0)(1.5,1.5)
\pspolygon[linewidth=.5pt,fillstyle=solid,fillcolor=tile](.75,0)(1.5,.75)(.75,1.5)(0,.75)(.75,0)
\pscircle[linewidth=1pt,fillstyle=solid,fillcolor=white](.75,0){.1}
\pscircle[linewidth=1pt,fillstyle=solid,fillcolor=white](1.5,.75){.1}
\pscircle[linewidth=1pt,fillstyle=solid,fillcolor=white](.75,1.5){.1}
\pscircle[linewidth=1pt,fillstyle=solid,fillcolor=white](0,.75){.1}
\end{pspicture}\;}

\def\tileb{\;
\begin{pspicture}[shift=-.65](0,0)(1.5,1.5)
\pspolygon[linewidth=.5pt,fillstyle=solid,fillcolor=tile](.75,0)(1.5,.75)(.75,1.5)(0,.75)(.75,0)
\pscircle[linewidth=1pt,fillstyle=solid,fillcolor=white](.75,0){.1}
\pscircle[linewidth=1pt,fillstyle=solid,fillcolor=white](1.5,.75){.1}
\pscircle[linewidth=1pt,fillstyle=solid,fillcolor=black](.75,1.5){.1}
\pscircle[linewidth=1pt,fillstyle=solid,fillcolor=white](0,.75){.1}
\end{pspicture}\;}

\def\tilec{\;
\begin{pspicture}[shift=-.65](0,0)(1.5,1.5)
\pspolygon[linewidth=.5pt,fillstyle=solid,fillcolor=tile](.75,0)(1.5,.75)(.75,1.5)(0,.75)(.75,0)
\pscircle[linewidth=1pt,fillstyle=solid,fillcolor=black](.75,0){.1}
\pscircle[linewidth=1pt,fillstyle=solid,fillcolor=white](1.5,.75){.1}
\pscircle[linewidth=1pt,fillstyle=solid,fillcolor=white](.75,1.5){.1}
\pscircle[linewidth=1pt,fillstyle=solid,fillcolor=white](0,.75){.1}
\end{pspicture}\;}

\def\tiled{\;
\begin{pspicture}[shift=-.65](0,0)(1.5,1.5)
\pspolygon[linewidth=.5pt,fillstyle=solid,fillcolor=tile](.75,0)(1.5,.75)(.75,1.5)(0,.75)(.75,0)
\pscircle[linewidth=1pt,fillstyle=solid,fillcolor=white](.75,0){.1}
\pscircle[linewidth=1pt,fillstyle=solid,fillcolor=white](1.5,.75){.1}
\pscircle[linewidth=1pt,fillstyle=solid,fillcolor=white](.75,1.5){.1}
\pscircle[linewidth=1pt,fillstyle=solid,fillcolor=black](0,.75){.1}
\end{pspicture}\;}

\def\tilee{\;
\begin{pspicture}[shift=-.65](0,0)(1.5,1.5)
\pspolygon[linewidth=.5pt,fillstyle=solid,fillcolor=tile](.75,0)(1.5,.75)(.75,1.5)(0,.75)(.75,0)
\pscircle[linewidth=1pt,fillstyle=solid,fillcolor=white](.75,0){.1}
\pscircle[linewidth=1pt,fillstyle=solid,fillcolor=black](1.5,.75){.1}
\pscircle[linewidth=1pt,fillstyle=solid,fillcolor=white](.75,1.5){.1}
\pscircle[linewidth=1pt,fillstyle=solid,fillcolor=white](0,.75){.1}
\end{pspicture}\;}

\def\tilef{\;
\begin{pspicture}[shift=-.65](0,0)(1.5,1.5)
\pspolygon[linewidth=.5pt,fillstyle=solid,fillcolor=tile](.75,0)(1.5,.75)(.75,1.5)(0,.75)(.75,0)
\pscircle[linewidth=1pt,fillstyle=solid,fillcolor=black](.75,0){.1}
\pscircle[linewidth=1pt,fillstyle=solid,fillcolor=white](1.5,.75){.1}
\pscircle[linewidth=1pt,fillstyle=solid,fillcolor=black](.75,1.5){.1}
\pscircle[linewidth=1pt,fillstyle=solid,fillcolor=white](0,.75){.1}
\end{pspicture}\;}

\def\tileg{\;
\begin{pspicture}[shift=-.65](0,0)(1.5,1.5)
\pspolygon[linewidth=.5pt,fillstyle=solid,fillcolor=tile](.75,0)(1.5,.75)(.75,1.5)(0,.75)(.75,0)
\pscircle[linewidth=1pt,fillstyle=solid,fillcolor=white](.75,0){.1}
\pscircle[linewidth=1pt,fillstyle=solid,fillcolor=black](1.5,.75){.1}
\pscircle[linewidth=1pt,fillstyle=solid,fillcolor=white](.75,1.5){.1}
\pscircle[linewidth=1pt,fillstyle=solid,fillcolor=black](0,.75){.1}
\end{pspicture}\;}

\def\tileD#1#2#3#4{\ \ 
\begin{pspicture}[shift=-.65](0,0)(1.5,1.5)
\pspolygon[linewidth=.5pt,fillstyle=solid,fillcolor=tile](.75,0)(1.5,.75)(.75,1.5)(0,.75)(.75,0)
\rput[t](.75,0){\scriptsize $#1$}
\rput[l](1.5,.75){\scriptsize $#2$}
\rput[b](.75,1.5){\scriptsize $#3$}
\rput[r](0,.75){\scriptsize $#4$}
\psline[linewidth=1pt,linestyle=dashed,linecolor=blue](.75,0)(.75,1.5)
\psline[linewidth=1pt,linestyle=dashed,linecolor=red](0,.75)(1.5,.75)
\end{pspicture}}

\def\tileF#1#2#3#4{\ \ 
\begin{pspicture}[shift=-.65](0,0)(1.5,1.5)
\pspolygon[linewidth=.5pt,fillstyle=solid,fillcolor=tile](.75,0)(1.5,.75)(.75,1.5)(0,.75)(.75,0)
\rput[t](.75,0){\scriptsize $#1$}
\rput[l](1.5,.75){\scriptsize $#2$}
\rput[b](.75,1.5){\scriptsize $#3$}
\rput[r](0,.75){\scriptsize $#4$}
\psline[linewidth=1pt,linestyle=dashed,linecolor=red](.75,0)(.75,1.5)
\psline[linewidth=1pt,linestyle=dashed,linecolor=blue](0,.75)(1.5,.75)
\end{pspicture}}

\def\tileX#1#2#3#4{\ \ 
\begin{pspicture}[shift=-.65](0,0)(1.5,1.5)
\pspolygon[linewidth=.5pt,fillstyle=solid,fillcolor=tile](.75,0)(1.5,.75)(.75,1.5)(0,.75)(.75,0)
\rput[t](.75,0){\scriptsize $#1$}
\rput[l](1.5,.75){\scriptsize $#2$}
\rput[b](.75,1.5){\scriptsize $#3$}
\rput[r](0,.75){\scriptsize $#4$}
\psline[linewidth=1pt,linestyle=dashed,linecolor=blue](.75,0)(.75,1.5)
\psline[linewidth=1pt,linestyle=dashed,linecolor=blue](0,.75)(1.5,.75)
\end{pspicture}}

\def\tileEj#1#2#3#4{\ \ 
\begin{pspicture}[shift=-.65](0,0)(1.5,1.5)
\pspolygon[linewidth=.5pt,fillstyle=solid,fillcolor=tile](.75,0)(1.5,.75)(.75,1.5)(0,.75)(.75,0)
\rput[t](.75,0){\scriptsize $#1$}
\rput[l](1.5,.75){\scriptsize $#2$}
\rput[b](.75,1.5){\scriptsize $#3$}
\rput[r](0,.75){\scriptsize $#4$}
\psline[linewidth=1pt,linestyle=dashed,linecolor=blue](0,.75)(1.5,.75)
\end{pspicture}}

\def\tileI#1#2#3#4{\ \ 
\begin{pspicture}[shift=-.65](0,0)(1.53,1.5)
\pspolygon[linewidth=.5pt,fillstyle=solid,fillcolor=tile](.75,0)(1.5,.75)(.75,1.5)(0,.75)(.75,0)
\rput[t](.75,0){\scriptsize $#1$}
\rput[l](1.5,.75){\scriptsize $#2$}
\rput[b](.75,1.5){\scriptsize $#3$}
\rput[r](0,.75){\scriptsize $#4$}
\psarc[linewidth=1.5pt,linecolor=blue](0,.75){.525}{-45}{45}
\psarc[linewidth=1.5pt,linecolor=blue](1.5,.75){.525}{135}{225}
\end{pspicture}}

\def\tileE#1#2#3#4{\ \ 
\begin{pspicture}[shift=-.65](0,0)(1.53,1.5)
\pspolygon[linewidth=.5pt,fillstyle=solid,fillcolor=tile](.75,0)(1.5,.75)(.75,1.5)(0,.75)(.75,0)
\rput[t](.75,0){\scriptsize $#1$}
\rput[l](1.5,.75){\scriptsize $#2$}
\rput[b](.75,1.5){\scriptsize $#3$}
\rput[r](0,.75){\scriptsize $#4$}
\psarc[linewidth=1.5pt,linecolor=blue](.75,0){.525}{45}{135}
\psarc[linewidth=1.5pt,linecolor=blue](.75,1.5){.525}{-135}{-45}
\end{pspicture}}

\nc{\Wtt}[6]{#1\!\!\left.\left(
\begin{matrix}
#5 & #4\\
#2 & #3
\end{matrix}
\right|#6\right)} 
\nc{\thf}{s} 
\nc{\lam}{\lambda}
\nc{\round}[1]{\left(#1\right)} 
\nc{\therm}{q}

\def\smbullet{\mbox{\tiny$\bullet$}}
\def\X{\Bbb X}

\begin{document}

\topmargin -5mm \oddsidemargin 5mm

\setcounter{page}{1}

\mbox{}\vspace{8mm}
\begin{center}
{\huge {\bf RSOS Quantum Chains Associated}}\\[6pt]
{\huge {\bf with Off-Critical Minimal Models}}\\[10pt]
{\huge {\bf and $\mathbb{Z}_n$ Parafermions}}

\vspace{12mm}
{\Large Davide Bianchini$^{\ast}$\!\!, Elisa Ercolessi$^\dagger$\!\!, Paul A. Pearce$^\ddag$, Francesco Ravanini$^\dagger$}\\[.5cm]
{\em {}$^\ast$Department of Mathematics, City University London}\\
{\em  Northampton Square, EC1V 0HB, UK}
\\[.2cm]
{\em {}$^\dagger$Department of Physics and Astronomy, University of Bologna and}\\
{\em  I.N.F.N. Sezione di Bologna, Via
Irnerio 46, 40126, Bologna, Italy}
\\[.2cm]
{\em {}$^\ddag$Department of Mathematics and Statistics, University of Melbourne}\\
{\em Parkville, Victoria 3010, Australia}
\\[.4cm]
{\tt Davide.Bianchini.1\,@\,city.ac.uk}
\quad
{\tt Elisa.Ercolessi\,@\,bo.infn.it}
\quad
{\tt P.Pearce\,@\,ms.unimelb.edu.au}
\quad
{\tt Francesco.Ravanini\,@\,bo.infn.it}

\end{center}

\vspace{15mm}
\centerline{{\bf{Abstract}}}
\vskip.3cm
\noindent
We consider the $\varphi_{1,3}$ off-critical perturbation ${\cal M}(m,m';t)$ of the general non-unitary minimal models where $2\le m\le m'$ and $m, m'$ are coprime and $t$ measures the departure from criticality corresponding to the $\varphi_{1,3}$
integrable perturbation. We view these models as the continuum scaling limit in the ferromagnetic Regime~III of the Forrester-Baxter Restricted Solid-On-Solid (RSOS) models on the square lattice. 
We also consider the RSOS models in the antiferromagnetic Regime~II related in the continuum scaling limit to $\mathbb{Z}_n$ parfermions with $n=m'-2$. Using an elliptic Yang-Baxter algebra of planar tiles encoding the allowed face configurations, we obtain the Hamiltonians of the associated quantum chains defined as the logarithmic derivative of the transfer matrices with periodic boundary conditions. The transfer matrices and Hamiltonians act on a vector space of paths on the $A_{m'-1}$ Dynkin diagram whose dimension is counted by generalized Fibonacci numbers.

\vspace{3mm}

\renewcommand{\thefootnote}{\arabic{footnote}}
\setcounter{footnote}{0}

\newpage
\tableofcontents

\newpage
\section{Introduction}

Since the pioneering work of Baxter~\cite{Baxter71,Baxter72} on commuting transfer matrices, it is well established that classical two-dimensional Yang-Baxter integrable statistical lattice models~\cite{BaxBook}  are deeply related to one-dimensional integrable quantum systems. Specifically, a one-dimensional quantum Hamiltonian ${\cal H}$ is obtained by taking the logarithmic derivative of the commuting transfer matrices $\vec T(u)$ in the limit $u\to 0$. This Hamiltonian limit applies to many interesting examples of integrable two-dimensional models with different symmetries, both at criticality and off-criticality, giving rise to various integrable one-dimensional quantum spin chains. Notable off-critical examples include the two-dimensional Ising model~\cite{Onsager} and XY spin chain~\cite{XY}, two-dimensional $\mathbb{Z}_n$ symmetric systems~\cite{FatZam85,ZnJMO} and their associated $\mathbb{Z}_n$ spin chains~\cite{VonGehlenRittenberg}, as well as the 8-vertex model and associated XYZ quantum spin chain considered by Baxter~\cite{Baxter71,Baxter72}.

The XY, XYZ and $\mathbb{Z}_n$ spin chains all have the property that the vector space of states ${\cal V}=\otimes^N \mathbb{C}^d$ factorizes over the sites, where $d=2$ or $n$ is the number of states per site. There are other interesting classes of quantum Hamiltonians, however, for which the space of states ${\cal V}$ does not factorize over the sites. In particular, the Restricted-Solid-On-Solid (RSOS) lattice models feature restrictions between neighbouring sites that are encoded by path spaces~\cite{PasquierDynkin} defined on Dynkin diagrams. At criticality, these lattice models are described~\cite{Pasquier} by the RSOS or height representation of the Temperley-Lieb algebra~\cite{TempLieb} and give rise to the Temperley-Lieb quantum chain in the Hamiltonian limit.
The simplest example at criticality is the golden chain~\cite{FeiguinEtAl} which has applications to anyons and topological quantum field theory. The golden chain is in fact the Hamiltonian limit of a critical two-dimensional model of interacting  hard squares~\cite{BaxP83} built on a vector space of paths on the $T_2 \equiv A_4 / \mathbb{Z}_2$ tadpole diagram.

In this paper, we are interested in a general formulation of ($\varphi_{1,3}$ perturbed) off-critical RSOS quantum chain Hamiltonians associated to the Yang-Baxter integrable two-dimensional $A$-type RSOS models of Andrews, Baxter and Forrester~\cite{ABF84,FB84}. Apart from their intrinsic interest, a recent motivation arises from the exact computation of entanglement entropy~\cite{SchumacherEtAl} of one-dimensional quantum systems. As an alternative to the well known methods~\cite{WilczekEtAl,CalabreseCardy,CardyCADoyon} of computing entanglement entropy, this quantity can be computed exactly~\cite{PeschelEtAl,FranchiniEtAl} on an infinite bipartite chain through an off-critical corner transfer matrix~\cite{CTM,BaxBook} approach. In particular, the calculation of entanglement entropy has been carried out~\cite{FranchiniDeLuca} for the unitary minimal models~\cite{ABF84,BPZ84} and generalized to nonunitary minimal models in \cite{BianchiniThesis}. These works, however, do not answer the physically relevant question as to precisely which one-dimensional off-critical quantum system it is whose entanglement entropy is computed.

In Section~2, we recall the concepts of planar algebras~\cite{Jones} and the diagrammatic multiplication of tiles by tensor contraction of indices (heights). We illustrate these concepts in the context of the off-critical golden chain and off-critical Yang-Lee chain interpreting the states as particle occupation with nearest-neighbour exclusion. In Section~3, we recall the elliptic Boltzmann face weights of the off-critical Forrester-Baxter RSOS models~\cite{FB84} and rewrite these as face transfer operators in terms of decorated planar algebra tiles. We also present sufficient algebraic relations to ensure the Yang-Baxter Equations (YBE) are satisfied. In Section 4, we use the tile operator formalism to compute the quantum Hamiltonian limit of the RSOS transfer matrix. We also show that specializing to criticality reproduces the usual quantum Temperley-Lieb chains. Relevant properties of the elliptic functions are summarized in Appendix~A.

\section{$T_2$ RSOS Quantum Chains}

\subsection{Off-critical golden and $\mathbb{Z}_3$ parafermionic chains}

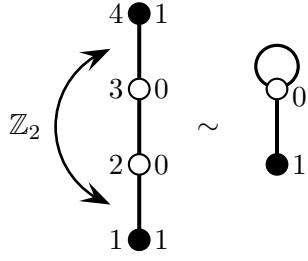
\begin{figure}[t]
\begin{center}
\begin{pspicture}[shift=-1.4](1,3)
\psline[linewidth=1.5pt](.5,0)(.5,3)
\pscircle*(.5,0){.15}
\pscircle*(.5,3){.15}
\pscircle[fillstyle=solid,fillcolor=white](.5,1){.15}
\pscircle[fillstyle=solid,fillcolor=white](.5,2){.15}
\rput[l](.1,0){$1$}
\rput[l](.1,1){$2$}
\rput[l](.1,2){$3$}
\rput[l](.1,3){$4$}
\rput[l](.7,0){$1$}
\rput[l](.7,1){$0$}
\rput[l](.7,2){$0$}
\rput[l](.7,3){$1$}
\psarc[linewidth=1pt,arrowsize=8pt]{<->}(.5,1.5){1.1}{110}{250}
\rput(-1.,1.5){$\mathbb{Z}_2$}
\end{pspicture}
$\ \sim\ $
\begin{pspicture}[shift=-1.4](1,3)
\psline[linewidth=1.5pt](.5,1)(.5,2)
\pscircle*(.5,1){.15}
\pscircle[linewidth=1.2pt](.5,2.3){.3}
\pscircle[fillstyle=solid,fillcolor=white](.5,2){.15}
\rput[l](.7,1){$1$}
\rput[l](.7,1.9){$0$}
\end{pspicture}
\end{center} 
\caption{\label{T2}The $\mathbb{Z}_2$ folding of the $A_4$ Dynkin diagram to the tadpole $T_2$.}
\end{figure}

The Interacting Hard Square (IHS) model~\cite{BaxBook,BaxP83} is defined on a square lattice. A site of the lattice is either occupied by a particle ($a=1$) or is vacant ($a=0$) subject to the nearest-neighbour exclusion $ab=0$ if $a,b=0,1$ are the occupancies of neighbouring sites. This adjacency condition is encoded in the $T_2=A_4/\mathbb{Z}_2$ tadpole diagram which is the $\mathbb{Z}_2$ folding of the $A_4$ Dynkin diagram as shown in Figure~\ref{T2}. A configuration along a periodic row of the square lattice is given by a path $\sigma=\{\sigma_1,\sigma_2,\ldots,\sigma_N\}$ where $\sigma_j=0,1$ and $\sigma_j\sigma_{j+1}=0$ for each $j=1,2,\ldots,N$ with $\sigma_{j+N}\equiv\sigma_j$ modulo $N$. 
The counting of paths from $a$ to $a$ in $N$ steps with $a=0,1$ is given by $F_{N+1}$ and $F_{N-1}$ respectively where the Fibonacci numbers are 
\bea
F_{n}=F_{n-1}+F_{n-2}=0,1,1,2,3,5,8,13,21,34,\ldots\qquad n=0,1,2,3,\ldots
\eea
In terms of the golden number $g=\half(1+\sqrt{5})$, the total number of periodic paths $L_N=F_{N+1}+F_{N-1}$ is given by the Lucas numbers
\bea
L_N=g^N+(-1)^N g^{-N}=1,3,4,7,11,18,29,47,\ldots\qquad N=1,2,3,\ldots
\eea

The $45^\circ$ rotated square lattice can be built, one elementary face at a time, using face transfer matrices. The IHS face transfer operators $\X_j(u)$ acting at position $j=1,2,\ldots,N$ are defined by
\psset{unit=.8cm}
\bea
\X_j(u)_\sigma^{\sigma'}=\prod_{k=1,2,\ldots,N\atop k\ne j} \delta(\sigma_k,\sigma'_k)\quad\  \diam {\sigma_j}{\sigma_{j+1}}{\sigma'_j}{\sigma_{j-1}}u
\eea
They act from the upper row configuration or path $\sigma'=\{\sigma'_1,\sigma'_2,\ldots,\sigma'_N\}$ to the lower row configuration or path $\sigma=\{\sigma_1,\sigma_2,\ldots,\sigma_N\}$. They act as the identity everywhere except in the column labelled by $j$ where they add a single face to the $45 ^\circ$ rotated square lattice. Multiplication of the face operators is by tensor contraction of indices given by the particle occupation numbers $\sigma_j=0,1$. As elements of the planar algebra~\cite{Jones}, the face operators decompose into a sum of seven elementary tiles
\begin{align}
\X_j(u)&=\!\!\!\diam{}{}{}{}u=\frac{s(2\lambda+u)}{s(2\lambda)}\tilea+\frac{s(2\lambda-u)}{s(2\lambda)}\tilef+\frac{s(\lambda-u)}{s(\lambda)}\Bigg(\tilee+\tiled\Bigg)
\nonumber\\[6pt]
&\ \ +\frac{s(\lambda+u)}{s(\lambda)}\tileg+\frac{s(u)}{\sqrt{s(\lambda)s(2\lambda)}}\Bigg(\tileb+\tilec\Bigg)\label{IHS}
\end{align}
where $s(u)=\vartheta_1(u)=\vartheta_1(u,\therm)$ is a standard elliptic theta
function~\cite{GR} of nome $q$, as in Appendix~A, $u$ is the spectral parameter and the crossing parameter is $\lambda=\frac{\pi}{5}$. Here we work in the $\mathbb{Z}_2$ folding of the $A_4$ face weights (\ref{Boltzmann1}) to (\ref{Boltzmann3}) in the symmetric gauge $g_a=\sqrt{s(a\lambda)}$ . The physical regimes of interest are
\bea
\begin{array}{rllcccc}
&\mbox{Regime~II:}\quad &0<q<1, \quad& -\lambda<u<0,&\ &\mbox{Hard hexagons}&\mbox{$\mathbb{Z}_3$ parafermions}\\
&\mbox{Regime~III:}\quad &0<q<1, \quad& \phantom{-} 0<u<\lambda,&\ &\mbox{Tricritical Ising}&\mbox{${\cal M}(4,5)$ minimal model}
\end{array}
\eea
We refer to the $u>0$ Regime~III as ferromagnetic and the $u<0$ Regime~II as antiferromagnetic. 
At criticality, $q=0$ and the central charges of the associated Conformal Field Theories (CFTs) are $c=\frac{4}{5}$ for hard hexagons (in the universality class of $\mathbb{Z}_3$ parafermions) and $c=\frac{7}{10}$ for the tricritical Ising model (unitary minimal model ${\cal M}(4,5)$).

Regarding the elementary tiles as operators acting on the upper row configuration to produce the lower row configuration, we write them respectively as
\bea
E_j=n_j^{000},n_j^{010},n_j^{001},n_j^{100},n_j^{101}, f_j,f_j^\dagger,\qquad 
n_j^{000}+n_j^{010}+n_j^{001}+n_j^{100}+n_j^{101}=I
\label{elemOps}
\eea
where $n_j^{abc}$ denotes a diagonal projector onto paths with $\sigma_{j-1}=a,\sigma_j=b,\sigma_{j+1}=c$. The non-diagonal operator $f_j\equiv f_j^{01}$ annihilates a particle at position $j$ whereas $f_{-j}\equiv f_j^\dagger\equiv f_j^{10}$ creates a particle at position $j$. These operators are fermionic in the sense that $f_j^2=f_{-j}^2=0$. However, they are not the modes of a free fermion since they implement nearest neighbour exclusion and satisfy the modified Canonical Anticommutation Relations (CAR)
\bea
\{f_j,f_k\}=(n_j^{000}+n_j^{010})\delta(j,-k)=n_j^{0\smbullet 0} \delta(j,-k)
\eea
where $n_j=n_j^{010}=f_j^\dagger f_j$ is the number operator giving the occupation of site $j$, $n_j^\dagger=n_j^{000}=f_j f_j^\dagger$ and $n_j^{0\smbullet 0}=n_j^{000}+n_j^{010}$ is the diagonal projector onto paths with $\sigma_{j-1}=\sigma_{j+1}=0$. Each of the 5 projectors can be writtem in terms of the fermion operators
\bea
n_j^{000}=f_j f_j^\dagger,\ n_j^{010}=f_j^\dagger f_j,\ n_j^{100}=f_{j-1}^\dagger f_j f_j^\dagger f_{j-1},\ n_j^{001}=f_{j+1}^\dagger f_j f_j^\dagger f_{j+1},\ n_j^{101}=f_{j-1}^\dagger f_{j-1} f_{j+1}^\dagger f_{j+1}
\eea

More generally, the fermionic modes $f_j$ generate an associative algebra ${\cal A}$. Acting on the vector space of paths from 0 to 0 in $N-1$ steps, the algebra ${\cal A}$ admits $F_N^2$ independent words $w$. 
These words map between any two arbitrary paths and take the canonical form
\bea
w=\prod_{j: \sigma_j=1}\!\! f_j^\dagger\; |0\rangle\langle0|\!\! \prod_{k: \sigma_k'=1}\!\! f_k=|\sigma\rangle\langle\sigma'|,\qquad |\sigma\rangle=\prod_{j: \sigma_j=1}\!\! f_j^\dagger\; |0\rangle
\eea
where $|0\rangle=|0;N\rangle$ denotes the vacuum path $\{\sigma_1,\sigma_2,\sigma_3,\ldots,\sigma_N\}=\{0,0,0,\ldots,0\}$ and the projector onto the vacuum state is 
\bea
|0\rangle\langle 0|=\prod_{j=2}^{N-1} n_j^{000}=\prod_{j=2}^{N-1} f_j f_j^\dagger
\eea
The matrix representatives of the words $w\in{\cal A}$ have entries which are all 0 or 1. The simplest way to work with this algebra is graphically in the planar algebra so that, for example, the relation $f_{j-1} n_j^{100}=n_j^{000}f_{j-1}=f_j f_j^\dagger f_{j-1}$ becomes
\bea
\raisebox{-.75cm}{
\begin{pspicture}(2.25,2.25)
\rput(.75,.75){\tileb}
\rput(1.5,1.5){\tiled}
\rput(.85,-.2){\tiny $j\!-\!1$}
\rput(1.6,.55){\tiny $j$}
\end{pspicture}}\ \ =\ \ 
\raisebox{-.75cm}{
\begin{pspicture}(2.25,2.25)
\rput(.75,1.5){\tileb}
\rput(1.5,.75){\tilea}
\rput(1.6,-.2){\tiny $j$}
\rput(.75,.45){\tiny $j\!-\!1$}
\end{pspicture}}\ \ =\ 
\raisebox{-1.3cm}{\begin{pspicture}(2.25,3.75)
\psline[linestyle=dashed](.83,.75)(.83,2.25)
\psline[linestyle=dashed](2.33,.75)(2.33,2.25)
\rput(.75,3){\tileb}
\rput(1.5,2.25){\tilec}
\rput(1.5,.75){\tileb}
\rput(.75,.4){\tiny $j\!-\!1$}
\rput(1.6,-.2){\tiny $j$}
\end{pspicture}}\ \ =\ |\{0,0,0,0\}\rangle\langle\{0,1,0,0\}|
\eea

Clearly, the elementary operators $E_j$ of (\ref{elemOps}) mutually commute if they are not at the same or adjacent sites
\bea
[E_j,E_k]=0,\qquad |j-k|>1
\eea
In addition, the elementary operators $E_j$ satisfy the single-site relations
\bea
&n_j^{abc}n_j^{a'b'c'}=\delta(a,a')\delta(b,b')\delta(c,c')\,n_j^{abc},\qquad
f_j^{ab}f_j^{b'a'}=\delta(b,b')f_j^{a,a'}&\\ 
&n_j^{abc} f_j^{b'b''}=\delta(a,0)\delta(b,b')\delta(c,0)\,f_j^{bb''},\qquad
f_j^{bb'}n_j^{ab''c}=\delta(a,0)\delta(b',b'')\delta(c,0)\,f_j^{bb''}&
\eea
where we set $f_j^{aa}=n_j^{0a0}$ for $a=0,1$. 
By using linearity and expanding out the 49 terms, the inversion relation
\bea
\X_j(u)\X_j(-u)=\frac{s(\lambda-u)s(\lambda+u)}{s(\lambda)^2}\,I,\qquad\qquad
\raisebox{-1.1cm}{\begin{pspicture}(1.5,3)
\psline[linestyle=dashed](.15,.75)(.15,2.25)
\psline[linestyle=dashed](1.65,.75)(1.65,2.25)
\rput(.75,.75){\diam ab{}du}
\rput(.75,2.25){\diam {\raisebox{-6pt}{\scriptsize $g$}}bcd{-u}}
\pscircle[fillstyle=solid,fillcolor=black](.9,1.5){.06}
\end{pspicture}}\ \ =\ \ \frac{s(\lambda-u)s(\lambda+u)}{s(\lambda)^2}\,\delta(a,c)
\label{InvReln}
\eea
follows immediately from the single-site relations and elliptic identities. This calculation is most easily carried out separately in the four blocks labelled by $0a0$, $100$, $001$, $101$.

The Yang-Baxter equation
\bea
\X_j(u)\X_{j+1}(u+v)\X_j(v)=\X_{j+1}(v)\X_j(u+v)\X_{j+1}(u),\qquad
\raisebox{-1.1cm}{\begin{pspicture}(0,0)(2.25,3)
\rput(.75,2.25){\diam {}{}{\,b'}av}
\rput(.75,.75){\diam bc{\raisebox{4pt}{\scriptsize $g$}}au}
\rput(1.5,1.5){\diam {}d{\raisebox{2pt}{\scriptsize $\,\,c'$}}{}{u\!+\!v}}
\psline[linestyle=dashed](.15,.75)(.15,2.25)
\pscircle[fillstyle=solid,fillcolor=black](.875,1.5){.06}
\end{pspicture}}\quad=\quad 
\raisebox{-1.1cm}{\begin{pspicture}(-.45,0)(1.5,3)
\rput(.75,2.25){\diam {}d{\,c'}{}u}
\rput(.75,.75){\diam cd{\raisebox{4pt}{\scriptsize $\,g'$}}{b\,}v}
\rput(0,1.5){\diam {}{}{b'}a{u\!+\!v}}
\psline[linestyle=dashed](1.65,.75)(1.65,2.25)
\pscircle[fillstyle=solid,fillcolor=black](.875,1.5){.06}
\end{pspicture}}
\label{YBE}
\eea
can similarly be established by considering separately the four blocks labelled by $0ab0$, $0a01$, $10a0$, $1001$ and using two-site relations combined with suitable elliptic identities. For example, in the $10a0$ block, the required two-site relations are
\bea
&n_j^{10a}f_{j+1}^{aa'}n_j^{10a'}=f_{j+1}^{a0}n_j^{100}f_{j+1}^{0a'}=f_{j+1}^{a1}n_j^{101}f_{j+1}^{1a'},\qquad a,a'=0,1
\eea
These relations, along with the obvious compatibility relations such as
\bea
n_j^{abc}n_{j+1}^{b'c'd}=\delta(b,b')\delta(c,c')n_j^{abc}n_{j+1}^{bcd}=\delta(b,b')\delta(c,c')n_{j+1}^{bcd}n_j^{ab'c'}
\eea
are straightforwardly read off from the diagrammatic product of elementary tiles. In this way, the single-site, two-site, and compatibility relations on the elementary tiles are sufficient to guarantee that the linear combination (\ref{IHS}) is a solution of the YBE. In this sense, the elementary operators $E_j$, subject to the single-site, two-site, compatibility and commutation relations, generate an off-critical associative Yang-Baxter algebra. 

For periodic boundary conditions, the off-critical quantum Hamiltonian associated with the IHS model is given by
\bea
{\cal H} =\mp\sum_{j=1}^N h_j,\qquad h_j=\frac{d}{du} \X_j(u)\Big|_{u=0}
\eea
where the upper sign is taken in Regime~III and the lower sign in Regime~II.
Explicitly, for the off-critical quantum chain, 
\begin{align}
h_j&=t_2\;\tilea-t_2\;\tilef-t_1\Bigg(\tilee+\tiled\Bigg)+t_1\;\tileg\nonumber\\[6pt]
&\ \ +\frac{\vartheta_1'(0)}{\sqrt{\vartheta_1(\lambda)\vartheta_1(2\lambda)}}\Bigg(\tileb+\tilec\Bigg),\qquad 
t_1=\frac{\vartheta_1'(\lambda)}{\vartheta_1(\lambda)},\ \; t_2=\frac{\vartheta_1'(2\lambda)}{\vartheta_1(2\lambda)}\label{IHShj}
\end{align}

\subsection{Critical golden and $\mathbb{Z}_3$ parafermionic chains}

At criticality $q=0$, $\frac{\vartheta_1(u)}{\vartheta_1(\lambda)}=\frac{\sin u}{\sin\lambda}$, $t_1=\cot\lambda$, $t_2=\cot 2\lambda$ and the Temperley-Lieb Hamiltonian of the periodic chain is
\bea
{\cal H}_{TL}=\begin{cases} -\disp\sum_{j=1}^N e_j,&\mbox{Regime~III (golden chain)}\\
+\disp\sum_{j=1}^N e_j,&\mbox{Regime~II ($\mathbb{Z}_3$ parafermions)}
\end{cases}
\eea
where, setting $\beta=2\cos\lambda=\frac{\sin 2\lambda}{\sin\lambda}$, the Temperley-Lieb generator is
\bea
e_j=\X_j(\lambda)=\frac{t_1 I+h_j}{t_1+t_2}=\tilea+\frac{1}{\beta}\;\tilef+\beta\;\tileg
+\frac{1}{\sqrt{\beta}}\Bigg(\tileb+\tilec\Bigg)
\eea
The properties of the Temperley-Lieb generators
\bea
e_j^2=\beta e_j,\qquad e_j e_{j\pm 1}e_j=e_j
\label{TLrelns}
\eea
follow from the algebraic properties of the fermionic operators $f_j$. In the ordered particle basis $\{\{1,0,1\}$, $\{0,0,1\}$, $\{1,0,0\}$, $\{0,1,0\}$, $\{0,0,0\}\}$, the matrix representation
\bea
\beta^{-1}\,e_j=\mbox{diag}(1,0,0)\oplus\begin{pmatrix} \beta^{-2}&\beta^{-3/2}\\ \beta^{-3/2}&\beta^{-1}\end{pmatrix}
\eea
agrees with the critical golden chain of \cite{FeiguinEtAl}. Off-criticality, the solution of the Yang-Baxter is built from the fermionic planar algebra and, at criticality, the Temperley-Lieb algebra is expressed in terms of the fermionic planar algebra. It is therefore clear that the fermionic planar algebra is more fundamental than the Temperley-Lieb algebra.

\subsection{Off-critical Yang-Lee chain}

The off-critical Yang-Lee model is the $\mathbb{Z}_2$ folding of the  $A_4$ RSOS model of \cite{FB84} with crossing parameter $\lambda=\frac{3\pi}{5}$. 
The face operators decompose into a sum of the same seven tiles as in (\ref{IHS})
\begin{align}
\X_j(u)&=\!\!\!\diam{}{}{}{}u=\frac{s(2\lambda+u)}{s(2\lambda)}\tilea+\frac{s(2\lambda-u)}{s(2\lambda)}\tilef+\frac{s(\lambda-u)}{s(\lambda)}\Bigg(\tilee+\tiled\Bigg)
\nonumber\\[6pt]
&\ \ +\frac{s(\lambda+u)}{s(\lambda)}\tileg+\frac{s(u)}{s(2\lambda)}\tileb+\frac{s(u)}{s(\lambda)}\tilec\label{YL}
\end{align} 
Because the model is nonunitary, we work with the $\mathbb{Z}_2$ folding of the $A_4$ RSOS model (\ref{Boltzmann1}) to (\ref{Boltzmann3}) with the non-symmetric gauge $g_a=1$. This choice breaks one of the diagonal reflection symmetries. 
The seven elementary tiles include the particle creation $f_j^\dagger$ and annihilation operators $f_j$. They satisfy the same algebraic relations as in the IHS model (both in the planar setting and as linear operators acting on the path basis) and implement nearest neighbour exclusion in the particle basis. 
From these algebraic relations, it follows that the face operators $\X_j(u)$ satisfy the inversion relation (\ref{InvReln}) with $\lambda=\frac{3\pi}{5}$ and the YBE (\ref{YBE}).
The relevant physical regime for the off-critical Yang-Lee model is
\bea
\begin{array}{rllcccc}
&\mbox{Regime~III:}\quad &0<q<1, \quad& \phantom{-} 0<u<\lambda,&\ &\mbox{Lee-Yang}&\mbox{${\cal M}(2,5)$ minimal model}
\end{array}
\eea
The continuum scaling limit at criticality with $q=0$ describes the nonunitary minimal CFT ${\cal M}(2,5)$ with central charge $c=-\frac{22}{5}$. Within the CFT, the relation between the particle basis and the Virasoro basis has been discussed in  \cite{FNO}.

For periodic boundary conditions, the off-critical quantum Hamiltonian associated with the Yang-Lee model is 
\begin{align}
{\cal H} =-\sum_{j=1}^N h_j,&\qquad h_j=t_2\;\tilea-t_2\;\tilef-t_1\Bigg(\tilee+\tiled\Bigg)+t_1\;\tileg\nonumber\\[6pt]
&\hspace{-.35in} +\frac{\vartheta_1'(0)}{\vartheta_1(2\lambda)}\;\tileb+\frac{\vartheta_1'(0)}{\vartheta_1(\lambda)}\;\tilec,\qquad 
t_1=\frac{\vartheta_1'(\lambda)}{\vartheta_1(\lambda)},\ \; t_2=\frac{\vartheta_1'(2\lambda)}{\vartheta_1(2\lambda)}\label{YLhj}
\end{align}

\subsection{Critical Yang-Lee chain}

At criticality $q=0$, $\frac{\vartheta_1(u)}{\vartheta_1(\lambda)}=\frac{\sin u}{\sin\lambda}$, $t_1=\cot\lambda$, $t_2=\cot 2\lambda$ and the Temperley-Lieb Hamiltonian of the periodic chain is
\bea
{\cal H}_{TL}=-\disp\sum_{j=1}^N e_j,&\qquad\mbox{Regime~III (Yang-Lee chain)}
\label{YLchain}
\eea
where, setting $\beta=2\cos\lambda=\frac{\sin 2\lambda}{\sin\lambda}$, the Temperley-Lieb generator is
\bea
e_j=\X_j(\lambda)=\frac{t_1 I+h_j}{t_1+t_2}=\tilea+\frac{1}{\beta}\;\tilef+\beta\;\tileg
+\frac{1}{\beta}\;\tileb+\tilec
\eea
Again, the properties of the Temperley-Lieb generators
\bea
e_j^2=\beta e_j,\qquad e_j e_{j\pm 1}e_j=e_j
\eea
follow from the algebraic properties of the fermionic operators $f_j$.

\section{Forrester-Baxter RSOS Models}
\subsection{Face weights, face operators and tiles}

\setlength{\unitlength}{24pt}
\psset{unit=24pt}
The minimal models ${\cal M}(m,m';t)$, where $t=q^2$ is a temperature-like variable, are described by the continuum scaling limit of the Forrester-Baxter Restricted Solid-On-Solid (RSOS) models~\cite{ABF84,FB84} on the square lattice. 
These models possess heights
{$a=1,2,\ldots,m'\!-\!1$} restricted so that nearest neighbour heights on the edges of the square lattice differ by $\pm 1$. The heights thus live on the $A_{m'-1}$ Dynkin diagram.
The Boltzmann weights are
\begin{align}
\omega_{1,a}(u)&=\Wtt{W}{a}{a\mp1}{a}{a\pm1}u\;=\;\faceu a{a\!\mp\!1}a{a\!\pm\!1}u\ \ 
\;=\;\frac{\thf(\lam-u)}{\thf(\lam)}\ \ \faceD {a}{a\!\mp\!1}{a}{a\!\pm\!1}
\label{Boltzmann1}\\
\omega_{2,a}^\pm(u)&=\Wtt{W}{a\mp1}{a}{a\pm1}{a}u\;=\;\faceu {a\!\mp\!1}a{a\!\pm\!1}au\ \ 
\;=\;{g_{a\mp 1}\over g_{a\pm 1}}\,
\frac{\thf((a\pm 1)\lam)}{\thf(a\lam)}\,
\frac{\thf(u)}{\thf(\lam)}\ \ \ \faceF {a\!\mp\!1}{a}{a\!\pm\!1}{a}
\label{Boltzmann2}\\
\omega_{3,a}^\pm(u)&=\Wtt{W}{a\pm1}{a}{a\pm1}{a}u\;=\;\faceu {a\!\pm\!1}a{a\!\pm\!1}au\ \ 
\;=\;
\frac{\thf(a\lam\pm u)}{\thf(a\lam)}\ \;\faceX {a\!\pm\!1}{a}{a\!\pm\!1}{a}
\label{Boltzmann3}
\end{align}
Here $\thf(u)=\vartheta_1(u,\therm)$ is a standard elliptic theta
function~\cite{GR}, as in Appendix~A, and 
$u$ is the spectral parameter. The crossing parameter is
\begin{equation}
\lambda={(m'-m)\pi\over m'},\qquad 2\le m<m',\qquad\mbox{$m,m'$ coprime}
\label{crossing}
\end{equation}
The elliptic nome is $\therm$, with $t=\therm^2$ measuring the
departure from criticality corresponding to the $\varphi_{1,3}$
integrable perturbation. 
The gauge factors $g_a$ are arbitrary functions of $a$. For convenience, we set $g_a=1$ but note that this choice breaks the $\mathbb{Z}_2$ refection symmetry about the SE-NW diagonal. The $\mathbb{Z}_2$ refection symmetry about the SW-NE diagonal is preserved. Setting $n=m'-2$, the physical regimes of interest are
\bea
\begin{array}{rllccc}
&\mbox{Regime~II:}\quad &0<q<1 \quad& -\lambda<u<0&\ &\mbox{$\mathbb{Z}_n$ parafermions}\\
&\mbox{Regime~III:}\quad &0<q<1 \quad& \phantom{-} 0<u<\lambda&\ &\mbox{${\cal M}(m,m')$ minimal model}
\end{array}
\eea

The tiles in (\ref{Boltzmann1}) to (\ref{Boltzmann3}) with dashed diagonals will be explained below. 
By abuse of notation, we regard such tiles as either scalars (matrix entries) or as operators (matrices) in a planar algebra. Fixing the direction of action as upwards, the faces give linear operators acting on a path space of heights. The RSOS face transfer operator $\X_j(u)$, acting at position $j=1,2,\ldots,N$, has entries
\bea
\X_j(u)_\sigma^{\sigma'}=\prod_{k=1,2,\ldots,N\atop k\ne j} \delta(\sigma_k,\sigma'_k)\quad\  \diam {\sigma_j}{\sigma_{j+1}}{\sigma'_j}{\sigma_{j-1}}u
\eea
The face operators act from the upper row configuration or path $\sigma'=\{\sigma'_1,\sigma'_2,\ldots,\sigma'_N\}$ to the lower row configuration or path $\sigma=\{\sigma_1,\sigma_2,\ldots,\sigma_N\}$ and add a single face to the $45^\circ$ rotated square lattice. The $\X_j(u)$ act as the identity (Kronecker delta) everywhere except on the column labelled by $j$. The row configurations with $\sigma_j, \sigma_j'\in A_{m'-1}$ respect the local adjacency rules so that $\sigma_{j+1}\!-\!\sigma_j=\pm 1$, etc. The boundary conditions can be open or periodic $\sigma_{j+N}\equiv \sigma_j$ by interpreting the sites $j$ modulo $N$. The number of paths for given boundary conditions is counted by generalized Fibonacci numbers.  
For periodic boundary conditions, the number of $N$-step paths is given by $\Tr A_{m'-1}^N$ where $A_{m'-1}$ denotes the adjacency matrix.

Following Jones~\cite{Jones}, it is sometimes convenient to adopt a diagrammatic planar algebra viewpoint. In our elliptic planar algebra, multiplication is implemented by local tensor contraction with the heights, denoted by $a,b,c$ or $\sigma_j$, as indices.
We thus decompose the face operators into a linear sum of elementary tiles with coefficients given by the elliptic Boltzmann weights
\begin{align}
\X_j(u)=\X_{a\,b\,c}^{a'b'c'}(u)&=\diam bc{\,b'}au\ =\omega_{1,b}(u)\,\delta(b,b')\bar\delta(a,c)\,\tile bc{\,b'}a\ 
+\omega_{2,a}^{b'\!\!-\!a}(u)\,\delta(a,c)\bar\delta(b,b')\,\tile bc{\,b'}a\nonumber\\[6pt]
&\qquad\qquad\qquad\qquad \mbox{}+\omega_{3,a}^{b'\!\!-\!a}(u)\,\delta(a,c)\delta(b,b')\,\tile bc{\,b'}a\\[6pt]
&=\omega_{1,b}(u)\,\tileD bc{\,b'}a\ 
+\omega_{2,a}^{b'\!\!-\!a}(u)\,\tileF bc{\,b'}a\ 
+\omega_{3,a}^{b'\!\!-\!a}(u)\,\tileX bc{\,b'}a
\end{align}
where we have introduced the complementary Kronecker delta
\bea
\bar\delta(a,b)=1-\delta(a,b)=\begin{cases}0, &a=b\\ 1,&a\ne b\end{cases}
\eea
A blue internal dashed line in a tile indicates a Kronecker delta and a red internal dashed line on a diagonal indicates a complementary Kronecker delta. 
For convenience, we have made the identifications $a=\sigma_{j-1}$, $b=\sigma_j$, $c=\sigma_{j+1}$ and $b'=\sigma'_j$. We will continue to make such identifications without further comment. 

\subsection{Inversion relation and YBE}

The Forrester-Baxter RSOS models are Yang-Baxter integrable~\cite{ABF84,FB84,BaxBook}. The face operators  satisfy the inversion and Yang-Baxter Equations (YBE)
\bea
\X_j(u)\X_j(-u)={s(\lambda-u)s(\lambda+u)\over s(\lambda)^2}\,I,\qquad \X_j(u)\X_{j+1}(u+v)\X_j(v)=\X_{j+1}(v)\X_j(u+v)\X_{j+1}(u)\\[-18pt]\nonumber
\eea
\bea
\raisebox{-1.1cm}{\begin{pspicture}(0,0)(1.5,3)
\rput(.75,2.25){\diam {}c{\,b'}a{-u}}
\rput(.75,.75){\diam bc{\raisebox{4pt}{\scriptsize $g$}}au}
\psline[linestyle=dashed](.15,.75)(.15,2.25)
\psline[linestyle=dashed](1.65,.75)(1.65,2.25)
\pscircle[fillstyle=solid,fillcolor=black](.875,1.5){.06}
\end{pspicture}}\quad =\ {s(\lambda-u)s(\lambda+u)\over s(\lambda)^2}\,\delta(b,b')\qquad\qquad\qquad
\raisebox{-1.1cm}{\begin{pspicture}(0,0)(2.25,3)
\rput(.75,2.25){\diam {}{}{\,b'}av}
\rput(.75,.75){\diam bc{\raisebox{4pt}{\scriptsize $g$}}au}
\rput(1.5,1.5){\diam {}d{\raisebox{2pt}{\scriptsize $\,\,c'$}}{}{u\!+\!v}}
\psline[linestyle=dashed](.15,.75)(.15,2.25)
\pscircle[fillstyle=solid,fillcolor=black](.875,1.5){.06}
\end{pspicture}}\quad=\quad 
\raisebox{-1.1cm}{\begin{pspicture}(-.45,0)(1.5,3)
\rput(.75,2.25){\diam {}d{\,c'}{}u}
\rput(.75,.75){\diam cd{\raisebox{4pt}{\scriptsize $\,g'$}}{b\,}v}
\rput(0,1.5){\diam {}{}{b'}a{u\!+\!v}}
\psline[linestyle=dashed](1.65,.75)(1.65,2.25)
\pscircle[fillstyle=solid,fillcolor=black](.875,1.5){.06}
\end{pspicture}}
\eea
The solid circle indicates that the internal heights $g,g'$ are summed over allowed values.

The elementary tile operators encode and impose the adjacency conditions of the allowed faces. In the height representation, given by action on configurations of rows of heights, the elementary tile operators $E_j$ are represented by binary matrices with entries
\bea
E_j=E_{a\,b\,c}^{a'b'c'}=\delta(a,a')\delta(c,c')\ \tile bc{\,b'}a\ \;=\;0,1
\eea
where $b=\sigma_j$ , $b'=\sigma_j'$ and so on. As linear operators, these matrices act as the identity everywhere except on the column labelled by $j$. 
The elementary tile operators $E_j$ satisfy a Yang-Baxter algebra in the form of a simple cell calculus which mimics the inversion relation and YBE
\bea
\raisebox{-1.1cm}{\begin{pspicture}(0,0)(1.5,3)
\rput(.75,2.25){\tile {}c{\,b'}a}
\rput(.75,.75){\tile bc{\raisebox{4pt}{\scriptsize $g$}}a}
\psline[linestyle=dashed](.15,.75)(.15,2.25)
\psline[linestyle=dashed](1.65,.75)(1.65,2.25)
\end{pspicture}}\quad =\ \tile bc{b'}a\qquad\quad
\raisebox{-1.1cm}{\begin{pspicture}(0,0)(2.25,3)
\rput(.75,2.25){\tile {}{}{\,b'}a}
\rput(.75,.75){\tile bc{\raisebox{4pt}{\scriptsize $g$}}a}
\rput(1.5,1.5){\tile {}d{\raisebox{2pt}{\scriptsize $\,\,c'$}}{}}
\psline[linestyle=dashed](.15,.75)(.15,2.25)
\end{pspicture}}\quad=\quad 
\raisebox{-1.1cm}{\begin{pspicture}(-.45,0)(1.5,3)
\rput(.75,2.25){\tile {}d{\,c'}{}}
\rput(.75,.75){\tile cd{\raisebox{4pt}{\scriptsize $\,g'$}}{b\,}}
\rput(0,1.5){\tile {}{}{b'}a}
\psline[linestyle=dashed](1.65,.75)(1.65,2.25)
\end{pspicture}}\qquad\quad E_jE_k=E_kE_j, \ \  |j-k|>1\label{tileRelns}
\eea
The principal difference is that there is no sum on the internal heights $g,g'$. These relations hold for all allowed elementary tile operators and all allowed fixed values of $g,g'$. The tile operators, with the defining relations (\ref{tileRelns}), replace the Pauli spin matrices $(\sigma^x_j,\sigma^y_j,\sigma^z_j)$ of the usual spin-\half quantum spin chains or the generators $e_j$ of the critical $q=0$ (trigonometric) Temperley-Lieb~\cite{TempLieb} chains.

The inversion relation involves linear operators on the single site $j$ acting on 2-step paths ($N=3$) whereas the YBE involves linear operators on the 2 sites $j,j\!+\!1$ acting on 3-step paths ($N=4$). Generically, these identities therefore reduce to identities on $4\times 4$ and $8\times 8$ matrices with the ordered bases ${\cal B}_N$ of paths
\def\+{\!+\!}
\def\-{\!-\!}
\begin{align}
{\cal B}_3&=\{(a,a\+1,a\+2),(a,a\+1,a),(a,a\-1,a),(a,a\-1,a\-2)\}\\[4pt]
{\cal B}_4&=\{(a,a\+1,a\+2,a\+3),(a,a\+1,a\+2,a\+1),(a,a\+1,a,a\+1),(a,a\-1,a,a\+1),\nonumber\\
&\qquad\quad(a,a\+1,a,a\-1),(a,a\-1,a,a\-1),(a,a\-1,a\-2,a\-1),(a,a\-1,a\-2,a\-3)\}
\end{align}
Explicitly, the $4\times 4$ and two $8\times 8$ matrices are
\begin{align}
\X_j(u)&=\begin{pmatrix}
\omega_{1,a}(u)&0&0&0\\
0&\omega_{3,a}^+(u)&\omega_{2,a}^-(u)&0\\
0&\omega_{2,a}^+(u)&\omega_{3,a}^-(u)&0\\
0&0&0&\omega_{1,a}(u)
\end{pmatrix}
\end{align}
\begin{align}
\X_j(u)&=\begin{pmatrix}
\omega_{1,a}(u) &0&0&0&0&0&0&0\\
0&\omega_{1,a}(u)&0&0&0&0&0&0\\
0&0& \omega_{3,a}^+(u)&\omega_{2,a}^-(u)&0&0&0&0\\
0&0&\omega_{2,a}^+(u)&\omega_{3,a}^-(u)&0&0&0&0\\
0&0&0&0&\omega_{3,a}^+(u)&\omega_{2,a}^-(u)&0&0\\
0&0&0&0&\omega_{2,a}^+(u)&\omega_{3,a}^-(u)&0&0\\
0&0&0&0&0&0&\omega_{1,a-1}(u)&0\\
0&0&0&0&0&0&0&\omega_{1,a-1}(u)
\end{pmatrix}
\end{align}
\begin{align}
\X_{j+1}(u)&=\begin{pmatrix}
\omega_{1,a+2}(u)&0&0&0&0&0&0&0\\
0&\omega_{3,a+1}^+(u)&\omega_{2,a+1}^-(u)&0&0&0&0&0\\
0&\omega_{2,a+1}^+(u)&\omega_{3,a+1}^-(u)&0&0&0&0&0\\
0&0&0&\omega_{1,a}(u)&0&0&0&0\\
0&0&0&0&\omega_{1,a}(u)&0&0&0\\
0&0&0&0&0&\omega_{3,a-1}^+(u)&\omega_{2,a-1}^-(u)&0\\
0&0&0&0&0&\omega_{2,a-1}^+(u)&\omega_{3,a-1}^-(u)&0\\
0&0&0&0&0&0&0&\omega_{1,a-2}(u)
\end{pmatrix}
\end{align}

The complete list of elementary tile identities used to prove the inversion relation and YBE are presented graphically and algebraically in Appendix~\ref{TileIds}. For generic $\lambda$, the required $\vartheta_1(u)$ elliptic identities are all special cases of the fundamental identity (\ref{FundId}) of Appendix~\ref{EllipticFns}.
Due to truncation of heights in the RSOS models, we need to remove some terms from the fundamental identities if their heights refer to configurations which are not allowed. Such restrictions enter the equations when $a=1,2,m'-2$ or $m'-1$. The elliptic identities still hold in such cases due to the properties
\bea
\vartheta_1(-u)=-\vartheta_1(u)=\vartheta(u+\pi),\qquad \vartheta_1(m'\lambda)=0
\eea

\subsection{Periodic row transfer matrices}

The periodic transfer matrix $\vec T(u)$ acting between the rows or paths $\sigma=\{\sigma_1,\sigma_2,\ldots,\sigma_N=\sigma_0\}$ and $\sigma'=\{\sigma'_1,\sigma'_2,\ldots,\sigma'_N=\sigma'_0\}$ has entries
\bea
\vec T(u)_\sigma^{\sigma'}=\prod_{j=1}^N X_{\sigma'_j,\sigma_{j},\sigma_{j+1}}^{\sigma'_j,\sigma'_{j+1},\sigma_{j+1}}(u)
\eea
The transfer matrix commutes with the shift operator $\Omega=\vec T(0)$ with entries
\bea
\psset{unit=1.cm}
\Omega_\sigma^{\sigma'}=\prod_{j=1}^N \delta(\sigma_j,\sigma'_{j+1})=\;
\raisebox{-.45cm}{\begin{pspicture}(0,0)(6,1)
\rput(.5,.5){\shifttile {\sigma_1}{\sigma_2}{\sigma_2'}{\sigma_1'}}
\rput(1.5,.5){\shifttile {}{\sigma_3}{\sigma_3'}{}}
\multirput(2.5,.5)(1,0){2}{\shifttile {}{}{}{}}
\rput(4.5,.5){\shifttile {\sigma_{N-1}}{\sigma_N}{\sigma_N'}{\sigma_{N-1}'}}
\rput(5.5,.5){\shifttile {}{\sigma_1}{\sigma_1'}{}}
\end{pspicture}}
\eea

\section{$A_{m'-1}$ RSOS Quantum Hamiltonians}

\subsection{Off-critical RSOS quantum Hamiltonians}
The quantum Hamiltonian $\cal H$ is given by the logarithmic derivative of the row transfer matrix at $u=0$
\bea
\vec T(u)\sim \Omega \exp(-u{\cal H})\sim\Omega[1-u{\cal H}+{\cal O}(u^2)],\qquad 
-{\cal H}={d\over du} \log \vec T(u)\big|_{u=0}={d\over du}\big[\Omega^{-1}\,\vec T(u)\big]_{u=0}
\eea
It follows that after shifting the energy by an amount ${s'(\lambda)\over s(\lambda)}$
\bea
-{\cal H}=\sum_{j=1}^N \Big[{s'(0)\over s(\lam)}\, F_j+X_j\Big],\qquad 
s'(0)=\vartheta'_1(0,q)=\vartheta_2(0,q)\vartheta_3(0,q)\vartheta_4(0,q)
\eea
where
\bea
\mbox{}\hspace{-12pt}
F_j\!=\!{s(\sigma'_j\lam)\over s(\sigma_{j+1}\lam)}\,\ \ \ \tileEj {\sigma_j}{\sigma_{j+1}}{\,\sigma'_j}{\sigma_{j-1}}\quad\ \ ,\ 
X_j\!=\!\Big[{s'(\lambda)\over s(\lambda)}
\!-\!{s'(0)\over s(\lam)}{s(\sigma'_j\lam)\over s(\sigma_{j+1}\lam)}
\!+\!(\sigma'_j\!-\!\sigma_{j+1}){s'(\sigma_{j+1}\lam)\over s(\sigma_{j+1}\lam)}\Big]
\quad\tileX {\sigma_j}{\sigma_{j+1}}{\,\sigma'_j}{\sigma_{j-1}}\quad
\eea
and $s'(u)=\vartheta'_1(u,q)$ denotes the derivative of $s(u)$ with respect to $u$. In this derivation we have made use of the identities
\medskip
\bea
\tileF {\sigma_j}{\sigma_{j+1}}{\sigma_j'}{\sigma_{j-1}}\quad \ \ +\quad\tileX {\sigma_j}{\sigma_{j+1}}{\sigma_j'}{\sigma_{j-1}}\quad\ \ =\quad\tileEj {\sigma_j}{\sigma_{j+1}}{\sigma_j'}{\sigma_{j-1}}\quad\ \ ,\quad\qquad
\tileD {\sigma_j}{\sigma_{j+1}}{\sigma_j'}{\sigma_{j-1}}\quad \ \ +\quad\tileX {\sigma_j}{\sigma_{j+1}}{\sigma_j'}{\sigma_{j-1}}\quad\ \ =\,I
\eea
which follow from the scalar relation $\delta(a,b)+\bar\delta(a,b)=1$.

\subsection{Critical RSOS quantum chains}
At criticality, with $q=0$ and $g_a=1$, the nonzero Boltzmann weights reduce to
\bea
\Wtt{W}abcdu
=\;{s(\lam\!-\!u)\over s(\lam)}\,\delta(a,c)
+{s(c\lam)\over s(b\lam)}\,{s(u)\over s(\lam)}\,\delta(b,d)\label{heightRep}
\eea
with $s(u)=\sin u$. The second term gives rise to matrix representations~\cite{OwczBaxt} of the Temperley-Lieb generators $e_j$ acting on paths built on the graph $A_{m'-1}$. The associated face operators are
\bea
\X_j(u)=\!\diam bc{\,b'}au\,={s(\lam\!-\!u)\over s(\lam)}\,I+{s(u)\over s(\lam)}\,e_j
={s(\lam\!-\!u)\over s(\lam)}\,\delta(b,b')\tileI bc{\,b'}a\ +\;{s(u)\over s(\lam)}\,\delta(a,c)\tileE bc{\,b'}a \label{critFaceOp}
\eea
where $I$ is the identity matrix and $j=1,2,\ldots, N$ fixes the position along a row. At criticality, the Temperley-Lieb generators coincide with the tiles $F_j$
\bea
e_j=\delta(a,c)\tileE bc{\,b'}a\ ={s(b'\lam)\over s(c\lam)}\,\tileEj bc{\,b'}a\ =F_j
\eea
and satisfy the Temperley-Lieb relations (\ref{TLrelns}) with the loop fugacity 
$\beta=2\cos\lambda=2\cos{(m'-m)\pi\over m'}$. The $X_j$ operators vanish at criticality since $\sigma_j'-\sigma_{j+1}=\pm 1$ and
\bea
\cot\lambda\pm\cot b\lambda=\frac{\sin(b\pm 1)\lambda}{\sin\lambda\sin b\lambda}
\eea

\section{Conclusion}

In this paper, we obtain the off-critical Hamiltonians of the periodic quantum chains corresponding to the RSOS lattice models in the anisotropic limit. 
This was achieved in both the ferromagnetic ($u>0$) Regime~III and antiferromagnetic ($u<0$) Regime~II. 
Off-criticality, for $t=q^2>0$, these regimes are associated with ${\Bbb Z}_2$-ordered and ${\Bbb Z}_{n}$-ordered phases respectively with $n=m'-2$. 
As the critical point is approached ($t,q\to 0$), the RSOS chains become conformally invariant and are described, in the continuum scaling limit, by CFTs. In Regime~III, the RSOS chains are associated with the unitary/nonunitary minimal models ${\cal M}(m,m')$ with $2\le m\le m'$ whereas, in Regime~II, the RSOS chains are associated with ${\Bbb Z}_{n}$ parafermions with $n=m'-2$.
We have thus found an off-critical quantum chain realization of the minimal models perturbed by the thermal operator $\varphi_{r,s}=\varphi_{1,3}$ with $\Delta_{1,3}=\frac{2m-m'}{m'}$ as well as a quantum chain realization of ${\Bbb Z}_n$ parafermions perturbed by the thermal operator $\varphi_{\ell,m}=\varphi_{2,0}$ with $\Delta_t=\Delta_{2,0}=\frac{2}{n+2}$. 
We stress that our analysis in Regime~III applies equally to nonunitary models and there is no need to restrict to unitary minimal models.
Of course, it is possible to write all of the RSOS Hamiltonians in terms of Pauli spin matrices. But, since this is not the natural language for these models, the resulting expressions are unwieldy and not illuminating.
At criticality, our quantum chains are associated with generalized interacting anyons in topological quantum liquids~\cite{FeiguinEtAl}.

Although we focused on the single row transfer matrix with periodic boundary conditions, the same critical  
or off-quantum chains
apply to the case of double row transfer matrices with simple (vacuum) open boundary conditions. 
Although the nonunitary RSOS lattice models admit negative Boltzmann weights and do not have a clear probabilistic interpretation, the associated quantum Hamiltonians and CFTs are well defined. 
The quantum Hamiltonians, being represented by sparse matrices, are much better suited for numerical calculations than the full double row transfer matrices.
Empirically, the numerical spectra of the double row transfer matrices and associated chains are real and the matrices are diagonalizable even though these matrices are not symmetric (Hermitian). This is probably explained by some PT symmetry.
A complete analysis of the open off-critical RSOS quantum chain looks interesting and is a possible 
direction for future research.
But let us just summarise the qualitative properties of the spectra of these open off-critical quantum chains.
The numerical eigenvalues are real and consist of a finite number ($2n$ in Regime~II, $2m-2$ in Regime III in the RSOS representation) of asymptotically degenerate lowest energy levels (ground states) separated by a finite gap to an order-$N$ band of next-lowest eigenvalues. As $t\to 0$, the gap vanishes and the expected conformal spectra of the ${\Bbb Z}_n$ parafermions/minimal models is reproduced.

To conclude, we mention that our results fill a gap in the reasoning of recent investigations into the exact calculation~\cite{FranchiniDeLuca,BianchiniThesis} of bipartite quantum entanglement in infinite chains through the corner transfer matrix approach~\cite{PeschelEtAl}.  
Corner transfer matrices~\cite{CTM} were introduced for {\em classical\/} off-critical two-dimensional lattice models. However, entanglement is a {\em quantum} property and refers to the associated one-dimensional quantum chain. It is precisely the Hamiltonian of this off-critical quantum chain that has been found and  described here. Now we know the correct identification of the models underlying the results of \cite{FranchiniDeLuca,BianchiniThesis}, it is possible, for example, to perform numerical DMRG calculations to explore entanglement properties more deeply by considering finite intervals, finite size, and so on. In particular, the observed behaviour~\cite{BianchiniThesis} of entanglement entropy growing as $\tfrac{1}{3}\,c_\text{\scriptsize eff}\log\xi$, where $\xi$ is the correlation length, provides the first independent  confirmation of general first principle results~\cite{DoyonEtAl} for the entanglement entropy of nonunitary conformal models. The precise physical meaning of these results for nonunitary theories poses further interesting questions that will stimulate future research.

\section*{Acknowledgments}
EE and FR thank the INFN grants QUANTUM and GAST for partial financial support. DB acknowledges the warm hospitality of the University of Bologna and INFN where part of this work was done. 
PAP is supported by the Melbourne Research Grant Support Scheme and gratefully acknowledges support during a visit to the Asia Pacific Centre for Theoretical Physics, Pohang, South Korea. We thank O. Castro-Alvaredo and B. Doyon for useful discussions.

\appendix

\section{Elliptic Functions}
\label{EllipticFns}

For convenience, in this section, we summarize the definitions of the elliptic functions and the fundamental elliptic identity used in this paper. The standard elliptic theta functions~\cite{GR} are
\begin{align}
\vartheta_1(u,q)&=2q^{1/4}\sin u
\prod_{n=1}^\infty (1-2q^{2n}\cos 2u+q^{4n})(1-q^{2n})\\
\vartheta_2(u,q)&=2q^{1/4}\cos u
\prod_{n=1}^\infty (1+2q^{2n}\cos 2u+q^{4n})(1-q^{2n})\\
\vartheta_3(u,q)&=\prod_{n=1}^\infty (1+2q^{2n-1}\cos
2u+q^{4n-2})(1-q^{2n})\\
\vartheta_4(u,q)&=\prod_{n=1}^\infty (1-2q^{2n-1}\cos
2u+q^{4n-2})(1-q^{2n})
\end{align}
Setting $s(u)=\vartheta_1(u,q)$, the fundamental elliptic identity as in (15.3.10) of \cite{BaxBook} is
\bea
s(u\!+\!v)s(u\!-\!v)s(x\!+\!y)s(x\!-\!y)=
s(u\!+\!x)s(u\!-\!x)s(v\!+\!y)s(v\!-\!y)\!-\!s(u\!+\!y)s(u\!-\!y)s(v\!+\!x)s(v\!-\!x)
\label{FundId}
\eea

\section{RSOS Elementary Tile Identities}
\label{TileIds}

The complete list of elementary tile identities used to prove the inversion relation and YBE for the generic RSOS models are presented graphically in the next two subsections. The tile identities are summarised algebraically in the subsequent subsection.

\subsection{Inversion relation tile identities}
\bea
\raisebox{-1.1cm}{\begin{pspicture}(0,0)(1.5,3)
\rput(.75,2.25){\tileD {}{a+2}{\,a\negg+\negg1}a}
\rput(.75,.75){\tileD {a\negg+\negg1}{a\negg+\negg2}{\rput(0.4,0){\scriptsize {a\negg+\negg1}}}a}
\psline[linestyle=dashed](.15,.75)(.15,2.25)
\psline[linestyle=dashed](1.65,.75)(1.65,2.25)
\end{pspicture}}\quad =\ \tileD {a\negg+\negg1}{a\negg+\negg2}{a\negg+\negg1}a\qquad\quad
\negg,\;\;\;\;\;
\raisebox{-1.1cm}{\begin{pspicture}(0,0)(1.5,3)
\rput(.75,2.25){\tileF {}{a}{\,{a\negg+\negg1}}a}
\rput(.75,.75){\tileF {a\negg+\negg1}{a}{\rput(0.4,0){\scriptsize {a\negg-\negg1}}}a}
\psline[linestyle=dashed](.15,.75)(.15,2.25)
\psline[linestyle=dashed](1.65,.75)(1.65,2.25)
\end{pspicture}}\quad \;\; =\ \tileX {a\negg+\negg1}{a}{a\negg+\negg1}a\qquad\quad
\negg,\;\;\;\;\;
\raisebox{-1.1cm}{\begin{pspicture}(0,0)(1.5,3)
\rput(.75,2.25){\tileX {}{a}{\,{a\negg+\negg1}}a}
\rput(.75,.75){\tileX {a\negg+\negg1}{a}{\rput(0.4,0){\scriptsize {a\negg+\negg1}}}a}
\psline[linestyle=dashed](.15,.75)(.15,2.25)
\psline[linestyle=dashed](1.65,.75)(1.65,2.25)
\end{pspicture}}\quad&=& \tileX {a\negg+\negg1}{a}{a\negg+\negg1}a\qquad\quad
\eea
\bea
\raisebox{-1.1cm}{\begin{pspicture}(0,0)(1.5,3)
\rput(.75,2.25){\tileF {}{a}{\,{a\negg+\negg1}}a}
\rput(.75,.75){\tileX {a\negg-\negg1}{a}{\rput(0.4,0){\scriptsize {a\negg-\negg1}}}a}
\psline[linestyle=dashed](.15,.75)(.15,2.25)
\psline[linestyle=dashed](1.65,.75)(1.65,2.25)
\end{pspicture}}\quad \;\; =\ \tileF {a\negg-\negg1}{a}{a\negg+\negg1}a\qquad\quad
\negg,\;\;\;\;\;
\;\;\raisebox{-1.1cm}{\begin{pspicture}(0,0)(1.5,3)
\rput(.75,2.25){\tileX {}{a}{\,{a\negg+\negg1}}a}
\rput(.75,.75){\tileF {a\negg-\negg1}{a}{\rput(0.4,0){\scriptsize {a\negg+\negg1}}}a}
\psline[linestyle=dashed](.15,.75)(.15,2.25)
\psline[linestyle=dashed](1.65,.75)(1.65,2.25)
\end{pspicture}}\quad = \tileF {a\negg-\negg1}{a}{a\negg+\negg1}a\qquad\quad
\negg,\;\;\;\;\;
\raisebox{-1.1cm}{\begin{pspicture}(0,0)(1.5,3)
\rput(.75,2.25){\tileX {}{a}{\,{a\negg-\negg1}}a}
\rput(.75,.75){\tileX {a\negg-\negg1}{a}{\rput(0.4,0){\scriptsize {a\negg-\negg1}}}a}
\psline[linestyle=dashed](.15,.75)(.15,2.25)
\psline[linestyle=dashed](1.65,.75)(1.65,2.25)
\end{pspicture}}\quad \;\; &=&\ \tileX {a\negg-\negg1}{a}{a\negg-\negg1}a\qquad\quad
\eea
\bea
\;\;\raisebox{-1.1cm}{\begin{pspicture}(0,0)(1.5,3)
\rput(.75,2.25){\tileF {}{a}{\,{a\negg-\negg1}}a}
\rput(.75,.75){\tileF {a\negg-\negg1}{a}{\rput(0.4,0){\scriptsize {a\negg+\negg1}}}a}
\psline[linestyle=dashed](.15,.75)(.15,2.25)
\psline[linestyle=dashed](1.65,.75)(1.65,2.25)
\end{pspicture}}\quad =\tileX {a\negg-\negg1}{a}{a\negg-\negg1}a\qquad\quad
\negg,\;\;\;\;\;
\raisebox{-1.1cm}{\begin{pspicture}(0,0)(1.5,3)
\rput(.75,2.25){\tileF {}{a}{\,{a\negg+\negg1}}a}
\rput(.75,.75){\tileX {a\negg-\negg1}{a}{\rput(0.4,0){\scriptsize {a\negg-\negg1}}}a}
\psline[linestyle=dashed](.15,.75)(.15,2.25)
\psline[linestyle=dashed](1.65,.75)(1.65,2.25)
\end{pspicture}}\quad \;\; &=&\ \tileF {a\negg-\negg1}{a}{a\negg+\negg1}a\qquad\quad
\negg,\;\;\;\;\;
\;\;\raisebox{-1.1cm}{\begin{pspicture}(0,0)(1.5,3)
\rput(.75,2.25){\tileX {}{a}{\,{a\negg+\negg1}}a}
\rput(.75,.75){\tileF {a\negg-\negg1}{a}{\rput(0.4,0){\scriptsize {a\negg+\negg1}}}a}
\psline[linestyle=dashed](.15,.75)(.15,2.25)
\psline[linestyle=dashed](1.65,.75)(1.65,2.25)
\end{pspicture}}\quad =  \tileF {a\negg-\negg1}{a}{a\negg+\negg1}a\qquad\quad
\eea
\bea
\negg\negg\raisebox{-1.1cm}{\begin{pspicture}(0,0)(1.5,3)
\rput(.75,2.25){\tileD {}{a\negg-\negg2}{\,a\negg-\negg1}a}
\rput(.75,.75){\tileD {a\negg-\negg1}{a\negg-\negg2}{\rput(0.4,0){\scriptsize {a\negg-\negg1}}}a}
\psline[linestyle=dashed](.15,.75)(.15,2.25)
\psline[linestyle=dashed](1.65,.75)(1.65,2.25)
\end{pspicture}}\quad &=&\ \tileD {a\negg-\negg1}{a\negg-\negg2}{a\negg-\negg1}a\qquad\quad
\eea

\subsection{YBE tile identities}
\bea
\raisebox{-1.1cm}{\begin{pspicture}(0,0)(2.25,3)
\rput(.75,2.25){\tileD {}{}{a\negg+\negg1}a}
\rput(.75,.75){\tileD {a\negg+\negg1}{}{}{a}}
\rput(1.5,1.5){\tileD {\rput(0.3,-0.1){a\negg+\negg2}}{a\negg+\negg3}{\rput(0.35,0.1){a\negg+\negg2}}{\rput(-0.4,0){a\negg+\negg1}}}
\psline[linestyle=dashed](.15,.75)(.15,2.25)
\end{pspicture}}\quad\quad=\quad\quad 
\raisebox{-1.1cm}{\begin{pspicture}(-.45,0)(1.5,3)
\rput(.75,2.25){\tileD {}{a\negg+\negg3}{a\negg+\negg2}{}}
\rput(.75,.75){\tileD {a\negg+\negg2}{a\negg+\negg3}{\rput(0.4,0){a\negg+\negg2}}{\rput(-0.35,-0.1){a\negg+\negg1}}}
\rput(0,1.5){\tileD {}{}{\rput(-0.35,0.1){a\negg+\negg1}}a}
\psline[linestyle=dashed](1.65,.75)(1.65,2.25)
\end{pspicture}}\qquad\quad
\negg,\;\;\;\;\;
\raisebox{-1.1cm}{\begin{pspicture}(0,0)(2.25,3)
\rput(.75,2.25){\tileD {}{}{a\negg+\negg1}a}
\rput(.75,.75){\tileD {a\negg+\negg1}{}{}{a}}
\rput(1.5,1.5){\tileX {\rput(0.3,-0.1){a\negg+\negg2}}{a\negg+\negg1}{\rput(0.35,0.1){a\negg+\negg2}}{\rput(-0.4,0){a\negg+\negg1}}}
\psline[linestyle=dashed](.15,.75)(.15,2.25)
\end{pspicture}}\quad\quad=\quad\quad 
\raisebox{-1.1cm}{\begin{pspicture}(-.45,0)(1.5,3)
\rput(.75,2.25){\tileF {}{a\negg+\negg1}{a\negg+\negg2}{}}
\rput(.75,.75){\tileF {a\negg+\negg2}{a\negg+\negg1}{\rput(0.2,0){a}}{\rput(-0.35,-0.1){a\negg+\negg1}}}
\rput(0,1.5){\tileX {}{}{\rput(-0.35,0.1){a\negg+\negg1}}a}
\psline[linestyle=dashed](1.65,.75)(1.65,2.25)
\end{pspicture}}\qquad\quad
\eea
\bea
\raisebox{-1.1cm}{\begin{pspicture}(0,0)(2.25,3)
\rput(.75,2.25){\tileX {}{}{a\negg+\negg1}a}
\rput(.75,.75){\tileD {a\negg+\negg1}{}{}{a}}
\rput(1.5,1.5){\tileF {\rput(0.3,-0.1){a\negg+\negg2}}{a\negg+\negg1}{\rput(0.35,0.1){a}}{\rput(-0.4,0){a\negg+\negg1}}}
\psline[linestyle=dashed](.15,.75)(.15,2.25)
\end{pspicture}}\quad\quad=\quad\quad 
\raisebox{-1.1cm}{\begin{pspicture}(-.45,0)(1.5,3)
\rput(.75,2.25){\tileX {}{a\negg+\negg1}{a}{}}
\rput(.75,.75){\tileF {a\negg+\negg2}{a\negg+\negg1}{\rput(0.4,0){a}}{\rput(-0.35,-0.1){a\negg+\negg1}}}
\rput(0,1.5){\tileX {}{}{\rput(-0.35,0.1){a\negg+\negg1}}a}
\psline[linestyle=dashed](1.65,.75)(1.65,2.25)
\end{pspicture}}\qquad\quad
\negg,\;\;\;\;\;
\raisebox{-1.1cm}{\begin{pspicture}(0,0)(2.25,3)
\rput(.75,2.25){\tileF {}{}{a\negg-\negg1}a}
\rput(.75,.75){\tileD {a\negg+\negg1}{}{}{a}}
\rput(1.5,1.5){\tileF {\rput(0.3,-0.1){a\negg+\negg2}}{a\negg+\negg1}{\rput(0.35,0.1){a}}{\rput(-0.4,0){a\negg+\negg1}}}
\psline[linestyle=dashed](.15,.75)(.15,2.25)
\end{pspicture}}\quad\quad=\quad\quad 
\raisebox{-1.1cm}{\begin{pspicture}(-.45,0)(1.5,3)
\rput(.75,2.25){\tileD {}{a\negg+\negg1}{a}{}}
\rput(.75,.75){\tileF {a\negg+\negg2}{a\negg+\negg1}{\rput(0.2,0){a}}{\rput(-0.35,-0.1){a\negg+\negg1}}}
\rput(0,1.5){\tileF {}{}{\rput(-0.35,0.1){a\negg-\negg1}}a}
\psline[linestyle=dashed](1.65,.75)(1.65,2.25)
\end{pspicture}}\qquad\quad
\eea
\bea
\raisebox{-1.1cm}{\begin{pspicture}(0,0)(2.25,3)
\rput(.75,2.25){\tileD {}{}{a\negg+\negg1}a}
\rput(.75,.75){\tileX {a\negg+\negg1}{}{}{a}}
\rput(1.5,1.5){\tileF {\rput(0.3,-0.1){a}}{a\negg+\negg1}{\rput(0.35,0.1){a\negg+\negg2}}{\rput(-0.4,0){a\negg+\negg1}}}
\psline[linestyle=dashed](.15,.75)(.15,2.25)
\end{pspicture}}\quad\quad=\quad\quad 
\raisebox{-1.1cm}{\begin{pspicture}(-.45,0)(1.5,3)
\rput(.75,2.25){\tileF {}{a\negg+\negg1}{a\negg+\negg2}{}}
\rput(.75,.75){\tileX {a}{a\negg+\negg1}{\rput(0.2,0){a}}{\rput(-0.35,-0.1){a\negg+\negg1}}}
\rput(0,1.5){\tileX {}{}{\rput(-0.35,0.1){a\negg+\negg1}}a}
\psline[linestyle=dashed](1.65,.75)(1.65,2.25)
\end{pspicture}}\qquad\quad
\negg,\;\;\;\;\;
\raisebox{-1.1cm}{\begin{pspicture}(0,0)(2.25,3)
\rput(.75,2.25){\tileD {}{}{a\negg+\negg1}a}
\rput(.75,.75){\tileX {a\negg+\negg1}{}{}{a}}
\rput(1.5,1.5){\tileF {\rput(0.3,-0.1){a}}{a\negg+\negg1}{\rput(0.35,0.1){a\negg+\negg2}}{\rput(-0.4,0){a\negg+\negg1}}}
\psline[linestyle=dashed](.15,.75)(.15,2.25)
\end{pspicture}}\quad\quad=\quad\quad 
\raisebox{-1.1cm}{\begin{pspicture}(-.45,0)(1.5,3)
\rput(.75,2.25){\tileX {}{a\negg+\negg1}{a\negg+\negg2}{}}
\rput(.75,.75){\tileF {a}{a\negg+\negg1}{\rput(0.4,0){a\negg+\negg2}}{\rput(-0.35,-0.1){a\negg+\negg1}}}
\rput(0,1.5){\tileD {}{}{\rput(-0.35,0.1){a\negg+\negg1}}a}
\psline[linestyle=dashed](1.65,.75)(1.65,2.25)
\end{pspicture}}\qquad\quad\\
\nonumber\\
\nonumber\\
\raisebox{-1.1cm}{\begin{pspicture}(0,0)(2.25,3)
\rput(.75,2.25){\tileF {}{}{a\negg+\negg1}a}
\rput(.75,.75){\tileF {a\negg+\negg1}{}{}{a}}
\rput(1.5,1.5){\tileD {\rput(0.3,-0.1){a}}{a\negg+\negg1}{\rput(0.35,0.1){a}}{\rput(-0.4,0){a\negg-\negg1}}}
\psline[linestyle=dashed](.15,.75)(.15,2.25)
\end{pspicture}}\quad\quad=\quad\quad 
\raisebox{-1.1cm}{\begin{pspicture}(-.45,0)(1.5,3)
\rput(.75,2.25){\tileX {}{a\negg+\negg1}{a}{}}
\rput(.75,.75){\tileX {a}{a\negg+\negg1}{\rput(0.4,0){a}}{\rput(-0.35,-0.1){a\negg+\negg1}}}
\rput(0,1.5){\tileX {}{}{\rput(-0.35,0.1){a\negg+\negg1}}a}
\psline[linestyle=dashed](1.65,.75)(1.65,2.25)
\end{pspicture}}\qquad\quad
\negg,\;\;\;\;\;
\raisebox{-1.1cm}{\begin{pspicture}(0,0)(2.25,3)
\rput(.75,2.25){\tileF {}{}{a\negg+\negg1}a}
\rput(.75,.75){\tileF {a\negg+\negg1}{}{}{a}}
\rput(1.5,1.5){\tileD {\rput(0.3,-0.1){a}}{a\negg+\negg1}{\rput(0.35,0.1){a}}{\rput(-0.4,0){a\negg-\negg1}}}
\psline[linestyle=dashed](.15,.75)(.15,2.25)
\end{pspicture}}\quad\quad=\quad\quad 
\raisebox{-1.1cm}{\begin{pspicture}(-.45,0)(1.5,3)
\rput(.75,2.25){\tileF {}{a\negg+\negg1}{a}{}}
\rput(.75,.75){\tileF {a}{a\negg+\negg1}{\rput(0.35,0){a\negg+\negg2}}{\rput(-0.35,-0.1){a\negg+\negg1}}}
\rput(0,1.5){\tileD {}{}{\rput(-0.35,0.1){a\negg+\negg1}}a}
\psline[linestyle=dashed](1.65,.75)(1.65,2.25)
\end{pspicture}}\qquad\quad
\eea
\bea
\raisebox{-1.1cm}{\begin{pspicture}(0,0)(2.25,3)
\rput(.75,2.25){\tileX {}{}{a\negg+\negg1}a}
\rput(.75,.75){\tileX {a\negg+\negg1}{}{}{a}}
\rput(1.5,1.5){\tileX {\rput(0.3,-0.1){a}}{a\negg+\negg1}{\rput(0.35,0.1){a}}{\rput(-0.4,0){a\negg+\negg1}}}
\psline[linestyle=dashed](.15,.75)(.15,2.25)
\end{pspicture}}\quad\quad=\quad\quad 
\raisebox{-1.1cm}{\begin{pspicture}(-.45,0)(1.5,3)
\rput(.75,2.25){\tileX {}{a\negg+\negg1}{a}{}}
\rput(.75,.75){\tileX {a}{a\negg+\negg1}{\rput(0.4,0){a}}{\rput(-0.35,-0.1){a\negg+\negg1}}}
\rput(0,1.5){\tileX {}{}{\rput(-0.35,0.1){a\negg+\negg1}}a}
\psline[linestyle=dashed](1.65,.75)(1.65,2.25)
\end{pspicture}}\qquad\quad
\negg,\;\;\;\;\;
\raisebox{-1.1cm}{\begin{pspicture}(0,0)(2.25,3)
\rput(.75,2.25){\tileX {}{}{a\negg+\negg1}a}
\rput(.75,.75){\tileX {a\negg+\negg1}{}{}{a}}
\rput(1.5,1.5){\tileX {\rput(0.3,-0.1){a}}{a\negg+\negg1}{\rput(0.35,0.1){a}}{\rput(-0.4,0){a\negg+\negg1}}}
\psline[linestyle=dashed](.15,.75)(.15,2.25)
\end{pspicture}}\quad\quad=\quad\quad 
\raisebox{-1.1cm}{\begin{pspicture}(-.45,0)(1.5,3)
\rput(.75,2.25){\tileF {}{a\negg+\negg1}{a}{}}
\rput(.75,.75){\tileF {a}{a\negg+\negg1}{\rput(0.35,0){a\negg+\negg2}}{\rput(-0.35,-0.1){a\negg+\negg1}}}
\rput(0,1.5){\tileD {}{}{\rput(-0.35,0.1){a\negg+\negg1}}a}
\psline[linestyle=dashed](1.65,.75)(1.65,2.25)
\end{pspicture}}\qquad\quad
\eea
\bea
\raisebox{-1.1cm}{\begin{pspicture}(0,0)(2.25,3)
\rput(.75,2.25){\tileX {}{}{a\negg-\negg1}a}
\rput(.75,.75){\tileF {a\negg+\negg1}{}{}{a}}
\rput(1.5,1.5){\tileD {\rput(0.3,-0.1){a}}{a\negg+\negg1}{\rput(0.35,0.1){a}}{\rput(-0.4,0){a\negg-\negg1}}}
\psline[linestyle=dashed](.15,.75)(.15,2.25)
\end{pspicture}}\quad\quad=\quad\quad 
\raisebox{-1.1cm}{\begin{pspicture}(-.45,0)(1.5,3)
\rput(.75,2.25){\tileD {}{a\negg+\negg1}{a}{}}
\rput(.75,.75){\tileX {a}{a\negg+\negg1}{\rput(0.4,0){a}}{\rput(-0.35,-0.1){a\negg+\negg1}}}
\rput(0,1.5){\tileF {}{}{\rput(-0.35,0.1){a\negg-\negg1}}a}
\psline[linestyle=dashed](1.65,.75)(1.65,2.25)
\end{pspicture}}\qquad\quad
\negg,\;\;\;\;\;
\raisebox{-1.1cm}{\begin{pspicture}(0,0)(2.25,3)
\rput(.75,2.25){\tileF {}{}{a\negg-\negg1}a}
\rput(.75,.75){\tileX {a\negg+\negg1}{}{}{a}}
\rput(1.5,1.5){\tileX {\rput(0.3,-0.1){a}}{a\negg+\negg1}{\rput(0.35,0.1){a}}{\rput(-0.4,0){a\negg+\negg1}}}
\psline[linestyle=dashed](.15,.75)(.15,2.25)
\end{pspicture}}\quad\quad=\quad\quad 
\raisebox{-1.1cm}{\begin{pspicture}(-.45,0)(1.5,3)
\rput(.75,2.25){\tileD {}{a\negg+\negg1}{a}{}}
\rput(.75,.75){\tileX {a}{a\negg+\negg1}{\rput(0.4,0){a}}{\rput(-0.35,-0.1){a\negg+\negg1}}}
\rput(0,1.5){\tileF {}{}{\rput(-0.35,0.1){a\negg-\negg1}}a}
\psline[linestyle=dashed](1.65,.75)(1.65,2.25)
\end{pspicture}}\qquad\quad
\eea
\bea
\raisebox{-1.1cm}{\begin{pspicture}(0,0)(2.25,3)
\rput(.75,2.25){\tileD {}{}{a\negg+\negg1}a}
\rput(.75,.75){\tileF {a\negg-\negg1}{}{}{a}}
\rput(1.5,1.5){\tileF {\rput(0.3,-0.1){a}}{a\negg+\negg1}{\rput(0.35,0.1){a\negg+\negg2}}{\rput(-0.4,0){a\negg+\negg1}}}
\psline[linestyle=dashed](.15,.75)(.15,2.25)
\end{pspicture}}\quad\quad=\quad\quad 
\raisebox{-1.1cm}{\begin{pspicture}(-.45,0)(1.5,3)
\rput(.75,2.25){\tileF {}{a\negg+\negg1}{a\negg+\negg2}{}}
\rput(.75,.75){\tileD {a}{a\negg+\negg1}{\rput(0.4,0){a}}{\rput(-0.35,-0.1){a\negg-\negg1}}}
\rput(0,1.5){\tileF {}{}{\rput(-0.35,0.1){a\negg+\negg1}}a}
\psline[linestyle=dashed](1.65,.75)(1.65,2.25)
\end{pspicture}}\qquad\quad
\negg,\;\;\;\;\;
\raisebox{-1.1cm}{\begin{pspicture}(0,0)(2.25,3)
\rput(.75,2.25){\tileF {}{}{a\negg+\negg1}a}
\rput(.75,.75){\tileX {a\negg-\negg1}{}{}{a}}
\rput(1.5,1.5){\tileD {\rput(0.3,-0.1){a}}{a\negg+\negg1}{\rput(0.35,0.1){a}}{\rput(-0.4,0){a\negg-\negg1}}}
\psline[linestyle=dashed](.15,.75)(.15,2.25)
\end{pspicture}}\quad\quad=\quad\quad 
\raisebox{-1.1cm}{\begin{pspicture}(-.45,0)(1.5,3)
\rput(.75,2.25){\tileX {}{a\negg+\negg1}{a}{}}
\rput(.75,.75){\tileD {a}{a\negg+\negg1}{\rput(0.4,0){a}}{\rput(-0.35,-0.1){a\negg-\negg1}}}
\rput(0,1.5){\tileF {}{}{\rput(-0.35,0.1){a\negg+\negg1}}a}
\psline[linestyle=dashed](1.65,.75)(1.65,2.25)
\end{pspicture}}\qquad\quad
\eea
\bea
\raisebox{-1.1cm}{\begin{pspicture}(0,0)(2.25,3)
\rput(.75,2.25){\tileX {}{}{a\negg+\negg1}a}
\rput(.75,.75){\tileF {a\negg-\negg1}{}{}{a}}
\rput(1.5,1.5){\tileX {\rput(0.3,-0.1){a}}{a\negg+\negg1}{\rput(0.35,0.1){a}}{\rput(-0.4,0){a\negg+\negg1}}}
\psline[linestyle=dashed](.15,.75)(.15,2.25)
\end{pspicture}}\quad\quad=\quad\quad 
\raisebox{-1.1cm}{\begin{pspicture}(-.45,0)(1.5,3)
\rput(.75,2.25){\tileX {}{a\negg+\negg1}{a}{}}
\rput(.75,.75){\tileD {a}{a\negg+\negg1}{\rput(0.4,0){a}}{\rput(-0.35,-0.1){a\negg-\negg1}}}
\rput(0,1.5){\tileF {}{}{\rput(-0.35,0.1){a\negg+\negg1}}a}
\psline[linestyle=dashed](1.65,.75)(1.65,2.25)
\end{pspicture}}\qquad\quad
\negg,\;\;\;\;\;
\raisebox{-1.1cm}{\begin{pspicture}(0,0)(2.25,3)
\rput(.75,2.25){\tileX {}{}{a\negg-\negg1}a}
\rput(.75,.75){\tileX {a\negg-\negg1}{}{}{a}}
\rput(1.5,1.5){\tileD {\rput(0.3,-0.1){a}}{a\negg+\negg1}{\rput(0.35,0.1){a}}{\rput(-0.4,0){a\negg-\negg1}}}
\psline[linestyle=dashed](.15,.75)(.15,2.25)
\end{pspicture}}\quad\quad=\quad\quad 
\raisebox{-1.1cm}{\begin{pspicture}(-.45,0)(1.5,3)
\rput(.75,2.25){\tileD {}{a\negg+\negg1}{a}{}}
\rput(.75,.75){\tileD {a}{a\negg+\negg1}{\rput(0.4,0){a}}{\rput(-0.35,-0.1){a\negg-\negg1}}}
\rput(0,1.5){\tileX {}{}{\rput(-0.35,0.1){a\negg-\negg1}}a}
\psline[linestyle=dashed](1.65,.75)(1.65,2.25)
\end{pspicture}}\qquad\quad
\eea
\bea
\raisebox{-1.1cm}{\begin{pspicture}(0,0)(2.25,3)
\rput(.75,2.25){\tileX {}{}{a\negg+\negg1}a}
\rput(.75,.75){\tileD {a\negg+\negg1}{}{}{a}}
\rput(1.5,1.5){\tileF {\rput(0.3,-0.1){a\negg+\negg2}}{a\negg+\negg1}{\rput(0.35,0.1){a}}{\rput(-0.4,0){a\negg+\negg1}}}
\psline[linestyle=dashed](.15,.75)(.15,2.25)
\end{pspicture}}\quad\quad=\quad\quad 
\raisebox{-1.1cm}{\begin{pspicture}(-.45,0)(1.5,3)
\rput(.75,2.25){\tileF {}{a\negg+\negg1}{a}{}}
\rput(.75,.75){\tileX {a\negg+\negg2}{a\negg+\negg1}{\rput(0.4,0){a\negg+\negg2}}{\rput(-0.35,-0.1){a\negg+\negg1}}}
\rput(0,1.5){\tileD {}{}{\rput(-0.35,0.1){a\negg+\negg1}}a}
\psline[linestyle=dashed](1.65,.75)(1.65,2.25)
\end{pspicture}}\qquad\quad
\negg,\;\;\;\;\;
\raisebox{-1.1cm}{\begin{pspicture}(0,0)(2.25,3)
\rput(.75,2.25){\tileF {}{}{a\negg-\negg1}a}
\rput(.75,.75){\tileF {a\negg-\negg1}{}{}{a}}
\rput(1.5,1.5){\tileX {\rput(0.3,-0.1){a}}{a\negg+\negg1}{\rput(0.35,0.1){a}}{\rput(-0.4,0){a\negg+\negg1}}}
\psline[linestyle=dashed](.15,.75)(.15,2.25)
\end{pspicture}}\quad\quad=\quad\quad 
\raisebox{-1.1cm}{\begin{pspicture}(-.45,0)(1.5,3)
\rput(.75,2.25){\tileD {}{a\negg+\negg1}{a}{}}
\rput(.75,.75){\tileD {a}{a\negg+\negg1}{\rput(0.4,0){a}}{\rput(-0.35,-0.1){a\negg-\negg1}}}
\rput(0,1.5){\tileX {}{}{\rput(-0.35,0.1){a\negg-\negg1}}a}
\psline[linestyle=dashed](1.65,.75)(1.65,2.25)
\end{pspicture}}\qquad\quad
\eea

\subsection{Algebraic form of tile identities}
The tile identities of the previous sections can all be written in a more compact form algebraically in terms of the elementary operators 
\bea
\nonumber\\
d_j = \quad\tileD{\sigma_j}{\sigma_{j+1}}{\sigma_j'}{\sigma_{j-1}}\qquad,\quad f_j =\quad \tileF{\sigma_j}{\sigma_{j+1}}{\sigma_j'}{\sigma_{j-1}}\qquad,\quad x_j = \quad\tileX{\sigma_j}{\sigma_{j+1}}{\sigma_j'}{\sigma_{j-1}}
\eea
Explicitly, the identities are
\bea
d_j^2 = d_j,\quad f_j^2=x_j,\quad x_j^2=x_j,\quad x_jf_j = f_j,\quad f_jx_j=f_j
\eea
\bea
\begin{array}{rclrclrcl}
f_jf_{j+1}f_j &\!\!\!=\!\!\!& f_{j+1}f_jf_{j+1},\qquad d_j x_{j+1} d_j &\!\!\!=\!\!\!& f_{j+1} x_j f_{j+1},\qquad d_j f_{j+1} x_j &\!\!\!=\!\!\!& f_{j+1} x_j x_{j+1}\\[3pt]
d_j f_{j+1}x_j &\!\!\!=\!\!\!& x_{j+1}d_j f_{j+1},\qquad d_j f_{j+1} d_j &\!\!\!=\!\!\!& f_{j+1} f_j d_{j+1},\qquad x_j f_{j+1} d_j &\!\!\!=\!\!\!& x_{j+1} x_j f_{j+1}\\[3pt]
x_j f_{j+1}d_j &\!\!\!=\!\!\!& f_{j+1}d_j x_{j+1},\qquad f_j d_{j+1} f_j &\!\!\!=\!\!\!& x_{j+1} x_j x_{j+1},\qquad f_j d_{j+1} f_j &\!\!\!=\!\!\!& f_{j+1} d_j f_{j+1}\\[3pt]
x_j x_{j+1}x_j &\!\!\!=\!\!\!& x_{j+1}x_j x_{j+1},\qquad x_j x_{j+1} x_j &\!\!\!=\!\!\!& f_{j+1} d_j f_{j+1},\qquad f_j d_{j+1} x_j &\!\!\!=\!\!\!& x_{j+1} f_j d_{j+1}\\[3pt]
x_j x_{j+1}f_j &\!\!\!=\!\!\!& x_{j+1}f_j d_{j+1},\qquad f_j f_{j+1} d_j &\!\!\!=\!\!\!& d_{j+1} f_j f_{j+1},\qquad x_j d_{j+1} f_j &\!\!\!=\!\!\!& d_{j+1} f_j x_{j+1}\\[3pt]
f_j x_{j+1}x_j &\!\!\!=\!\!\!& d_{j+1}f_j x_{j+1},\qquad x_j d_{j+1} x_j &\!\!\!=\!\!\!& d_{j+1} x_j d_{j+1},\qquad f_j x_{j+1} f_j &\!\!\!=\!\!\!& d_{j+1} x_j d_{j+1}
\end{array}
\eea

The elementary operators $d_j$ and $x_j$ are diagonal operators. In contrast, the operators $f_j$ with $\sigma_j>\sigma_j'$ are height raising while the operators $f_{-j}=f_j^\dagger$ with $\sigma_j<\sigma_j'$ are height lowering operators. The operators $f_j$ are fermionic in the sense that $f_j^2=0$. All states $|\sigma\rangle$ can be obtained by acting with the height raising operators on the vacuum path $|0\rangle=\{1,2,1,2,\ldots,2,1\}$. As before, it follows that all words $|\sigma\rangle\langle\sigma|$ in the algebra can be constructed as monomials in $f_j$ and $f_j^\dagger$ combined with the projector $|0\rangle\langle0|$ onto the vacuum state.

\end{document}